\newcommand*{\ATLASLATEXPATH}{}
\documentclass[cernpreprint,UKenglish,texlive=2011,cover,PAPER]{atlasdoc}
\pdfoutput=1

\usepackage[biblatex=false]{\ATLASLATEXPATH atlaspackage}

\usepackage{\ATLASLATEXPATH atlascontribute}

\usepackage{\ATLASLATEXPATH atlasphysics}

\graphicspath{{logos/}{figures/}}

\hypersetup{pdftitle={ATLAS draft},pdfauthor={The ATLAS Collaboration}}

\AtlasTitle{Measurement of $D^{*\pm}$, $D^\pm$ and $D_s^\pm$ meson production cross sections in $pp$ collisions at $\sqrt{s}=7\,$TeV with the ATLAS detector}

\author{The ATLAS Collaboration}

\date{\today}

\AtlasRefCode{BPHY-2013-04}

\PreprintIdNumber{CERN-PH-EP-2015-288}

\AtlasJournal{Nucl.\ Phys.\ B}

\AtlasAbstract{%
The production of $\dspm$, $\dcpm$ and $\dsspm$ charmed mesons
has been measured
with the ATLAS detector in $pp$ collisions at $\sqrt{s}=7\,$TeV
at the LHC,
using
data corresponding to
an integrated luminosity of $280\,$nb$^{-1}$.
The charmed mesons have been reconstructed in the range
of transverse momentum $3.5<\pt(D)<100\gev$
and pseudorapidity $|\eta(D)|<2.1$.
The differential cross sections as a function
of transverse momentum and pseudorapidity
were measured for $\dspm$ and $\dcpm$ production.
The next-to-leading-order QCD predictions are
consistent with the data
in the visible kinematic region 
within the large theoretical uncertainties.
Using
the visible $D$ cross sections
and an extrapolation to the full kinematic phase space,
the strangeness-suppression factor
in charm fragmentation,
the fraction of charged non-strange $D$ mesons
produced in a vector state,
and the total cross section of charm production
at $\sqrt{s}=7\,$TeV
were derived.
}

\AtlasCoverSupportingNote{Supporting note}{https://cds.cern.ch/record/1666889}
\AtlasCoverCommentsDeadline{XX XX 2015}

\AtlasCoverAnalysisTeam{Leonid Gladilin, Miriam Watson}

\AtlasCoverEdBoardMember{Ian~Brock~(chair)}
\AtlasCoverEdBoardMember{Chariclia~Petridou}
\AtlasCoverEdBoardMember{Dario~Barberis}
\AtlasCoverEdBoardMember{Sergey~Burdin}

\AtlasCoverEgroupEditors{atlas-BPHY-2013-04-editors@cern.ch}

\AtlasCoverEgroupEdBoard{atlas-BPHY-2013-04-editorial-board@cern.ch}

\usepackage{cite}

\newcommand{\dsp}        {\mbox{$D^{\ast +}$}}
\newcommand{\dspm}       {\mbox{$D^{\ast \pm}$}}
\newcommand{\dc}         {\mbox{$D^+$}}
\newcommand{\dcpm}       {\mbox{$D^{\pm}$}}
\newcommand{\dssp}       {\mbox{$D_s^+$}}
\newcommand{\dsspm}      {\mbox{$D_s^{\pm}$}}
\newcommand{\dz}         {\mbox{$D^{0}$}}
\newcommand{\br}         {\mbox{${\cal B}_{D^{\ast +}\rightarrow D^0 \pi^+}$}}
\newcommand{\acc}        {\mbox{$\cal A$}}
\newcommand{\bran}       {\mbox{$\cal B$}}
\newcommand{\lum}        {\mbox{$\cal L$}}
\newcommand{\sig}        {\mbox{$\sigma$}}
\newcommand{\fcds}       {\mbox{$f(c \rightarrow D^{\ast +})$}}
\newcommand{\fcdc}       {\mbox{$f(c \rightarrow D^+)$}}
\newcommand{\fcdss}      {\mbox{$f(c \rightarrow D_s^+)$}}
\newcommand{\sv}         {\mbox{$\sigma^{\text{vis}}$}}
\newcommand{\scct}       {\mbox{$\sigma_{c{\bar c}}^{\text{tot}}$}}
\newcommand{\dr}         {\mbox{${\text{d}}$}}
\newcommand{\mub}        {\mbox{${\muup{\text{b}}}$}}
\begin{document}

\maketitle

\tableofcontents

\section{\label{secIntro}Introduction}

Measurements of heavy-quark production at
the Large Hadron Collider (LHC) 
provide a means to test
perturbative
quantum chromodynamics (QCD)
calculations
at the highest available collision energies.
Since the current calculations suffer from large theoretical uncertainties,
the experimental constraints on heavy-quark production cross sections
are important
for measurements in the electroweak and Higgs sectors,
and in searches for new physics phenomena,
for which heavy-quark production
is often an important background process.

Charmed mesons are produced in the hadronisation
of charm and bottom quarks,
which are copiously produced
in $pp$ collisions at $\sqrt{s}=7\,$TeV.
The ATLAS detector\footnote
{The ATLAS coordinate system is a Cartesian right-handed system,
with the coordinate origin at the nominal interaction point.
The anti-clockwise beam direction defines the positive $z$-axis,
with the $x$-axis pointing to the centre of the LHC ring.
Polar ($\theta$) and azimuthal ($\phi$) angles are measured
with respect to this reference system,
which corresponds
to the centre-of-mass frame of the colliding protons.
The pseudorapidity is defined as $\eta=-\ln\tan(\theta/2)$
and the transverse momentum is defined as $\pt=p\sin\theta$.
The rapidity is defined as $y=0.5\ln((E+p_{z})/(E-p_{z}))$,
where $E$ and $p_{z}$ refer to energy and longitudinal momentum,
respectively.}~\cite{ATLASPerformanceBook}
at the LHC
has been used previously to measure $\dsp$ mesons\footnote
{Hereafter, charge conjugation is implied.}
produced in jets~\cite{ATLASDsjets} and
in bottom hadron decays
in association
with muons~\cite{ATLASDsmuon}.
Associated production of $D$ mesons
and $W$ bosons has been also studied by the ATLAS collaboration~\cite{ATLASWc}.
Production of $D$ mesons
in the hadronisation of charm quarks
has been studied by the ALICE collaboration in the central
rapidity range
($|y|<0.5$)~\cite{ALICE7Ds,ALICE7Dss}
and by the LHCb collaboration at forward rapidities ($2.0<y<4.5$)~\cite{LHCb7Ds}.
Open-charm production was also measured by the CDF collaboration~\cite{CDF196D}
at the Tevatron collider
in $p{\bar p}$ collisions at $\sqrt{s}=1.96\,$TeV.

In this paper,
measurements of the inclusive
$\dsp$, $\dc$ and $\dssp$
production cross sections
and their comparison with next-to-leading-order (NLO) QCD
calculations
are presented.
Contributions from both
charm hadronisation and
bottom hadron decays have been included
in the measured visible $D$ production cross sections
and in the NLO QCD predictions.
The measured visible cross sections
have been extrapolated
to the cross sections for $D$ meson production
in charm hadronisation
in the full kinematic phase space,
after subtraction of the cross-section fractions
originating from bottom production.
The extrapolated cross sections
have been used to calculate
the total cross section of charm production
in $pp$ collisions at $\sqrt{s}=7\,$TeV
and two fragmentation ratios
for charged charmed mesons:
the strangeness-suppression factor
and the fraction of charged non-strange $D$ mesons
produced in a vector state.

\section{\label{secDetector}The ATLAS detector}

A detailed description of the ATLAS detector can be found
elsewhere~\cite{ATLASPerformanceBook}. 
A brief outline of the components most relevant to this analysis
is given below.

The ATLAS inner detector has full coverage in $\phi$,
covers the pseudorapidity range $|\eta|<2.5$
and operates inside an axial magnetic field of $2\,$T
of a superconducting solenoid.
It consists of a silicon pixel
detector (Pixel),
a silicon microstrip detector (semiconductor tracker, SCT) and
a transition radiation tracker (TRT).
The inner-detector barrel (end-cap) parts consist of
3 ($2\times3$) Pixel layers,
4 ($2\times9$) double-layers of single-sided SCT strips
and 73 ($2\times160$) layers of TRT straws.
The TRT straws enable track-following up to $|\eta|=2.0$.

The calorimeter system is placed outside the solenoid.
A high-resolution liquid-argon electromagnetic sampling calorimeter
covers the pseudorapidity range $|\eta|<3.2$.
This calorimeter is complemented by hadronic calorimeters, built using
scintillating tiles in the range $|\eta|<1.7$ and liquid-argon
technology in
the end-cap ($1.5<|\eta|<3.2$). Forward calorimeters extend the coverage
to $|\eta|<4.9$.

The ATLAS detector has a three-level trigger system~\cite{ATLAStrigger}.
For the measurement of $D$ mesons with $3.5<\pt<20\gev$ (low-$\pt$ range),
two complementary minimum-bias triggers
are used.
The first trigger relies on the first-level
trigger signals from
the Minimum Bias Trigger Scintillators (MBTS).
The MBTS are mounted at each end of the inner detector in front
of the liquid-argon end-cap calorimeter cryostats at $z=\pm3.56\,$m
and are segmented into eight sectors in azimuth
and two rings in pseudorapidity ($2.09<|\eta|<2.82$ and $2.82<|\eta|<3.84$).
The MBTS trigger used in this analysis
is configured to require at least one hit above
threshold.
The second minimum-bias trigger
uses
the inner detector at the second-level trigger
to select inelastic events
on randomly chosen bunch crossings (Random).
For $D$ mesons with $20<\pt<100\gev$ (high-$\pt$ range),
the first-level calorimeter-based jet triggers are used.
The jet triggers use coarse detector information
to identify areas in the calorimeter
with energy deposits above certain thresholds.
A simplified jet-finding algorithm based on a sliding window
of configurable size is used to trigger events. The algorithm
uses towers with a granularity of
$\Delta\phi\times\Delta\eta=0.2\times 0.2$ as inputs.
In this paper, the first-level jet triggers with
energy thresholds of 5, 10 and $15\gev$
are used. No further jet selection requirements are applied
at the second and third trigger levels.

The integrated luminosity is calculated by measuring
interaction rates using several ATLAS devices at small angles to
the beam direction,
with the absolute calibration obtained from beam-separation scans.
The uncertainty of the luminosity measurement
for the event sample used in this analysis
is estimated
to be $3.5\%$~\cite{ATLASlumi}.

\section{\label{secMC}Event simulation}

To model inelastic events produced in $pp$ collisions,
a large sample of Monte Carlo (MC) simulated events is prepared using
the PYTHIA~6.4~\cite{pythia6} MC generator.
The simulation is performed using leading-order matrix elements
for all $2\rightarrow 2$ QCD processes.
Initial- and final-state parton showering is used
to simulate the effect of higher-order processes.
The MRST LO*~\cite{MRSTLOstar}
parameterisation is used for the parton distribution functions (PDF)
of the proton.
The charm- and bottom-quark masses are set
to $1.5\,$GeV and $4.8\,$GeV, respectively.
The event sample is generated using
the ATLAS AMBT1 set of tuned parameters~\cite{ATLAStracks}.
The fraction of the $D$ meson sample
produced in bottom-hadron decays
($\sim$$10\%$)
is normalised
using the measured production cross section
of $b$-hadrons decaying
to $\dsp\mu^-X$ final states~\cite{ATLASDsmuon}.
 
The generated events are passed through
a full ATLAS detector simulation~\cite{ATLASGEANT4}
based on GEANT4~\cite{GEANT4a, GEANT4b}
and processed with the same reconstruction program as used for the data.

\section{\label{secNLO}QCD calculations}
\label{sec-xsec}

The measured $D$ cross sections are compared with
the fixed-order next-to-leading-logarithm (FONLL)~\cite{fonll1,fonll2,fonll3}
predictions, with the
general-mass variable-flavour-number scheme (GM-VFNS)~\cite{gmvfns1,gmvfns2,gmvfns3}
calculations and with the
NLO QCD calculations matched with a leading-logarithm parton-shower
MC simulation (NLO-MC).
A web interface was used to obtain
up-to-date FONLL predictions~\cite{fonllweb},
while the GM-VFNS predictions have been provided by
their authors.
Two methods are presently available
for performing the NLO-MC matched calculations:
MC@NLO~\cite{mcatnloMETHOD} and POWHEG~\cite{powhegMETHOD}.
Their implementations
in the codes MC@NLO~3.42~\cite{mcatnloHQ} and POWHEG-hvq~1.01~\cite{powhegHQ}
are used.
MC@NLO~3.42 is matched with the HERWIG~6.5~\cite{herwig} MC event generator,
while POWHEG-hvq~1.01 is used with both HERWIG~6.5 and PYTHIA~6.4.

The main differences between the GM-VFNS and the other calculations considered
here
originate from differences between the so-called massless and massive schemes.
In the massive scheme, the heavy quark $Q$ appears only in the final state
and the ${m_Q^2}/{p_{{\rm T},Q}^2}$ power terms of the perturbative series
are correctly accounted for,
where $p_{{\rm T},Q}$ is the transverse momentum of the heavy quark
and $m_Q$ is its pole mass.
The massive-scheme calculations are not reliable for $p_{{\rm T},Q}\gg m_Q$
due to neglected terms of the type $\ln(p_{{\rm T},Q}^2/m_Q^2)$.
In the massless scheme, the heavy quark occurs as an initial-state parton
and the large logarithmic terms are absorbed into
the heavy-quark contribution to the proton PDF,
and into the fragmentation functions of the heavy-quark transition to a hadron.
The massless calculations are reliable only for $p_{{\rm T},Q}\gg m_Q$
due to the assumption that $m_Q=0$.
The FONLL and GM-VFNS calculations were developed
to obtain reliable predictions for $p_{{\rm T},Q}\approx m_Q$.
In FONLL, the massive and massless predictions are matched exactly
up to $\mathcal O(\alpha_s^3)$,
and spurious higher-order terms with potentially unphysical behaviour
are damped using a weighting function.
The FONLL parton cross sections are convolved with
non-perturbative fragmentation functions.
GM-VFNS combines the massless predictions with the massive
${m_Q^2}/{p_{{\rm T},Q}^2}$ power terms and derives subtraction terms
by comparing the massive and massless cross sections
in the limit $m_Q\rightarrow 0$.
The large logarithmic terms
in GM-VFNS remain absorbed in the PDF and in
perturbatively evolved fragmentation functions with
a non-perturbative input.
Unlike other calculations, GM-VFNS considers fragmentation
to $D$ mesons from light quarks and gluons
in addition to the heavy-quark fragmentation~\cite{gmvfns_ff}.

All predictions
are obtained
using the CTEQ6.6~\cite{CTEQ66}
parameterisation for the proton PDF.
The value of the QCD coupling constant
is set to $\alpha_s(m_Z)=0.118$
in accord with the central CTEQ6.6 analysis.
Both the charm and bottom contributions
to the charmed meson production cross sections
are included in all predictions.
The charm-quark pole mass is set
to $1.5\,$GeV in all calculations.
The bottom-quark pole mass is set
to $4.75\,$GeV in the FONLL, MC@NLO and POWHEG
calculations. In the GM-VFNS calculations,
the bottom-quark pole mass is set
to $4.5\,$GeV.
The renormalisation and factorisation scales are set to
$\mu_r=\mu_f=\mu$, where $\mu$ is defined as
$$\mu^2=m_Q^2+p_{{\rm T},Q}^2~$$
in the FONLL and GM-VFNS calculations.
For MC@NLO,
$$\mu^2=m_Q^2+\frac{(p_{{\rm T},Q}+p_{{\rm T},{\bar Q}})^2}{4}\,,$$
where $p_{{\rm T},Q}$ and $p_{{\rm T},{\bar Q}}$ are the transverse momenta of the produced
heavy quark and antiquark, respectively, and $m_Q$ is the heavy-quark pole mass.
For POWHEG,
$$\mu^2=m_Q^2+(m^2_{Q{\bar Q}}/4-m_Q^2)\cdot\sin^2(\theta_Q)\,,$$
where $m_{Q{\bar Q}}$ is the invariant mass of
the produced $Q{\bar Q}$ system
and $\theta_Q$ is the polar angle of the heavy quark in
the $Q{\bar Q}$ system centre-of-mass frame.

The specific FONLL fragmentation functions~\cite{fonllweb,fonll_ff}
as well as the GM-VFNS fragmentation functions~\cite{gmvfns_ff}
were obtained using $e^+e^-$ data.
In the case of the NLO-MC matched calculations,
the heavy-quark hadronisation is performed
using the cluster model~\cite{HerwigCluster} when interfaced to HERWIG.
When interfaced to PYTHIA, the Lund string model~\cite{LundModel} with
the Bowler modification~\cite{Bowler} of the Lund symmetric fragmentation
function~\cite{LundFunction} for heavy quarks is used.

In the FONLL, MC@NLO and POWHEG calculations,
the fragmentation fractions
of heavy quarks hadronising as a particular
charmed meson, $f(Q\rightarrow D)$,
are set to experimental values
obtained by averaging the LEP
measurements
in hadronic $Z$ decays~\cite{ctod}.
They are summarised in Table~\ref{tab:ftod}.
In GM-VFNS,
the fragmentation fractions of heavy quarks, light quarks and gluons
were obtained
using $e^+e^-$ data,
along
with the fragmentation functions~\cite{gmvfns_ff}.

\begin{table}[hbt!]
\begin{center}
\renewcommand{\arraystretch}{1.4}
\begin{tabular}{c|c} \hline
& LEP data \\
\hline
$\fcds$ & $0.236 \pm 0.006 \pm 0.003$ \\
$\fcdc$ & $0.225 \pm 0.010 \pm 0.005$  \\
$\fcdss$ & $0.092 \pm 0.008 \pm 0.005$  \\
\hline
$f(b\rightarrow D^{*\pm})$ & $0.221 \pm 0.009 \pm 0.003$ \\
$f(b\rightarrow D^{\pm})$ & $0.223 \pm 0.011 \pm 0.005$  \\
$f(b\rightarrow D_s^{\pm})$ & $0.138 \pm 0.009 \pm 0.006$  \\
\hline
\end{tabular}
\caption{
The fractions of $c$ and $b$ quarks hadronising as a particular charmed meson,
$f(Q \rightarrow D)$,
obtained by averaging the LEP measurements~\cite{ctod}.
The first uncertainties are the combined statistical and systematic
uncertainties of the measurements.
The second uncertainties originate from
uncertainties
in the relevant branching fractions.
}
\label{tab:ftod}
\end{center}
\end{table}

The following sources of theoretical uncertainty
are considered for the FONLL, MC@NLO and POWHEG predictions:
\begin{itemize}
\item{Scale uncertainty. The uncertainty was determined by
varying $\mu_r$ and $\mu_f$ independently to $\mu/2$ and $2\mu$,
with the additional constraint $1/2<\mu_r/\mu_f<2$,
and selecting the largest positive and negative variations.
}
\item{Pole-mass uncertainty. The uncertainty is determined by
varying the charm- and bottom-quark masses independently
by $\pm0.2\,$GeV and $\pm0.25\,$GeV, respectively.
The total $m_Q$ uncertainty is obtained by adding
in quadrature
separately the positive
and negative cross-section variations.
}
\item{PDF uncertainty. The uncertainty is determined by
using the CTEQ6.6 PDF error eigenvectors.
For MC@NLO and POWHEG, the PDF $\alpha_s$ uncertainties
are also calculated using eigenvectors for $\pm0.002$ variations
of $\alpha_s$.
Following the PDF4LHC recommendations~\cite{pdf4lhc},
the CTEQ6.6 PDF and PDF $\alpha_s$ uncertainties,
provided at 90$\%$ confidence level (CL),
are scaled
to 68$\%$ CL.
The total PDF uncertainty (for FONLL)
or the combined PDF and $\alpha_s$ (PDF$\oplus\alpha_s$)
uncertainty (for MC@NLO and POWHEG)
is obtained by adding
in quadrature
separately the positive
and negative cross-section variations.
}
\item{Fragmentation-fraction uncertainty.
The uncertainty is
the combined statistical and systematic
uncertainty of the LEP measurements~\cite{ctod}.
The uncertainties on the fragmentation fractions originating from
uncertainties
in the relevant branching fractions
are not included
because they affect experimental and theoretical cross-section calculations
in the same way and can be ignored in the comparison.
}
\end{itemize}

For the POWHEG+PYTHIA predictions,
the hadronisation uncertainty
for each $D$ meson is obtained
as a sum in quadrature of the corresponding fragmentation-fraction
uncertainty and  the fragmentation-function uncertainty.
The latter uncertainty is determined
by using the Peterson fragmentation function~\cite{Peterson}
with extreme choices~\cite{Chrin,CacciariFF,NasonFF,OPAL_FF,ZEUS_FF}
of the fragmentation parameter:
$0.02$ and $0.1$ for charm fragmentation,
and $0.002$ and $0.01$ for bottom fragmentation.

Only the scale uncertainty, which is dominant, is calculated for GM-VFNS
by varying three scale parameters: the renormalisation scale, the factorisation
scale for initial-state singularities and the factorisation scale
for final-state singularities.
These three scales are varied independently to $\mu/2$ and $2\mu$,
with the additional constraint for the ratio of any two scales to be
between $1/2$ and $2$,
and the largest positive and negative variations are selected.

\section{\label{secSelection}Event selection}

The data used in this analysis
were collected in 2010
with the ATLAS detector
in $pp$ collisions at $\sqrt{s}=7\tev$
at the LHC.
The crossing angle of the colliding protons
was either zero or negligible
in the rapidity range of the measurement.
To measure $D$ mesons with $\pt<20\gev$,
the events collected with the minimum-bias
MBTS and
Random
triggers are used;
these triggers
are unbiased
for the events of interest~\cite{ATLAStrigger}.
However, the rate from the triggers exceeded the allotted trigger bandwidth
after the initial data-taking period
and thus prescale factors were applied to reduce the output rate.
Taking into account the prescale factors, the data sample 
corresponds to an integrated luminosity of $1.04\,$nb$^{-1}$.
To measure $D$ mesons in the intervals $20<\pt<30\gev$,
$30<\pt<40\gev$ and $40<\pt<100\gev$,
the first-level jet triggers with
energy thresholds of 5, 10 and $15\gev$, respectively,
are used.
The trigger efficiencies for the corresponding $D$ meson $\pt$
ranges are above $90\%$.
The efficiencies are derived from the MC simulation.
The simulation 
uncertainties
are estimated from data--MC comparisons using independent trigger selections
with softer thresholds on the jet energy or energy in the
electromagnetic calorimeter. 
The triggers with energy thresholds of 5 and $10\gev$ were prescaled
during some parts of the data-taking period;
their corresponding integrated luminosities are
$28\,$nb$^{-1}$ and $90\,$nb$^{-1}$,
respectively.
The data sample taken with the unprescaled jet trigger
with the energy threshold of $15\gev$
corresponds to  an integrated luminosity of $280\,$nb$^{-1}$.

The event samples are processed using the standard offline ATLAS detector
calibration and event reconstruction~\cite{ATLASPerformanceBook,ATLASIDCalib}.
Only events with
at least three reconstructed tracks
with $\pt>100\mev$
and at least one reconstructed primary-vertex
candidate~\cite{ATLASVxRec}
are kept for the reconstruction of charmed mesons.

\section{\label{secReconstruction}Reconstruction of charmed mesons}

The $\dsp$, $~\dc$ and $~\dssp~$
charmed mesons
are reconstructed in
the range of transverse momentum $3.5<\pt(D)<100\gev$
and pseudorapidity $|\eta(D)|<2.1$.
As no significant differences between results
for positively and negatively charged charmed mesons
are observed, all results are presented for the combined samples.

Charmed meson candidates are reconstructed using tracks measured
in the inner tracking detector.
To ensure high reconstruction efficiency and
good momentum resolution,
each track is required to satisfy
$|\eta|<2.5$, have
at least one hit in the Pixel detector and at least four hits in the SCT.
The $\dr E/\dr x$ particle identification
with the Pixel detector~\cite{ATLASperformance}
is not used since it is not effective in the kinematic ranges utilised
for the charmed-meson reconstruction.

There can be several primary-vertex candidates in an event
due to multiple collisions per bunch crossing.
To identify the heavy-quark production vertex,
requirements on the  $D$ meson transverse impact parameter,
$d_0$, and longitudinal impact parameter,
$z_0$,
with respect to the primary-vertex candidate
are imposed.
In the rare case
($<1\%$)
that more than one vertex
satisfies these requirements,
the hard-scatter primary vertex is taken to be the one
with the largest sum of the squared transverse
momenta of its associated tracks.

For $D$ mesons with momenta in the low-$\pt$ range,
the background from non-signal track combinations
(combinatorial background) is significantly reduced by requiring
$\pt(D^{*+}, D^+, D_s^+)/ \sum \pt(\rm{track}) > 0.05$,
where $\sum p_{\rm{T}}(\rm{track})$ is
the scalar sum of the transverse momenta of all tracks
associated with the primary vertex.
MC studies indicate that due to properties of heavy-quark
fragmentation,
more than $99\%$ of $D$ signals
satisfy this selection criterion.
Further background rejection is achieved by imposing requirements on
the $D^0$ (from the $D^{*+}\rightarrow D^0\pi^+$ decay),
$D^+$ and $D^+_s$ transverse decay lengths\footnote
{The transverse decay length of a particle is the transverse distance between
the primary or production vertex and the particle decay vertex, projected along
the transverse momentum of the particle.}
with respect to the primary vertex,
$L_{xy}$,
and on the transverse momenta and decay angles
of the charmed meson decay products.
The requirement values are tuned using
the MC simulation
to enhance signal-to-background ratios
while keeping acceptances high.

The details of the
reconstruction for each of the three charmed meson samples
are given in the next subsections.

\subsection{Reconstruction of $\dsp$ mesons}
\label{sec-recds}

The $\dsp$ mesons are identified using the decay
$\dsp\rightarrow\dz\pi^{+}_{s}\rightarrow(K^{-}\pi^{+})\pi^{+}_{s}$.
The pion from the $\dsp\rightarrow D^0\pi^+$ decay is referred to as
the ``soft'' pion, $\pi^+_s$,
because its momentum is limited
by the small mass difference between the $\dsp$ and $\dz$.

In each event, pairs of tracks from oppositely charged particles,
each with $\pt>1\gev$,
are combined to form $\dz$ candidates.
Any additional track, with $\pt>0.25\gev$,
is combined with the $\dz$ candidate
to form a $\dsp$ candidate.
The three tracks of the $\dsp$ candidate
are fitted using a constraint on the
$\dsp\rightarrow\dz\pi^{+}_{s}\rightarrow(K^{-}\pi^{+})\pi^{+}_{s}$
topology, i.e.
the two tracks of the $\dz$ candidate are required
to intersect at a single vertex and the $\dz$
trajectory is required to intersect with the third track,
producing the $\dsp$ vertex.
To calculate the $\dz$ candidate invariant mass, $m(K \pi)$,
kaon and pion masses are assumed in turn for each track.
The additional track is assigned the pion mass and
this pion is
required to have
a charge opposite to that of the kaon.
The mass $m(K \pi)$, the three-particle invariant mass $m(K \pi\pi_s)$,
and the mass difference, $\Delta m=m(K \pi \pi_s)-m(K \pi)$,
are calculated using
the track momenta refitted to the decay topology.
To suppress combinatorial background
the following requirements are used:
\begin{itemize}
\item{
$\chi^2<25$,
where $\chi^2$ is the $\dsp$ candidate fit quality.
The requirement value is loose as the signal-to-background ratio
decreases rather slowly with $\chi^2$.
}
\item{
$|d_0(\dsp)|<0.5\,$mm.
}
\item{
$|z_0(\dsp)\sin\theta(\dsp)|<0.5\,$mm.
}
\item{
$L_{xy}(\dz)>0.1\,$mm.
}
\item{
$|\cos\theta^*(K)|<0.95$,
where $\theta^*(K)$
is the angle between the kaon in the $K\pi$ rest frame and the $K\pi$
line of flight in the laboratory frame.
}
\end{itemize}

\begin{figure}[]
  \includegraphics[width=0.95\textwidth]{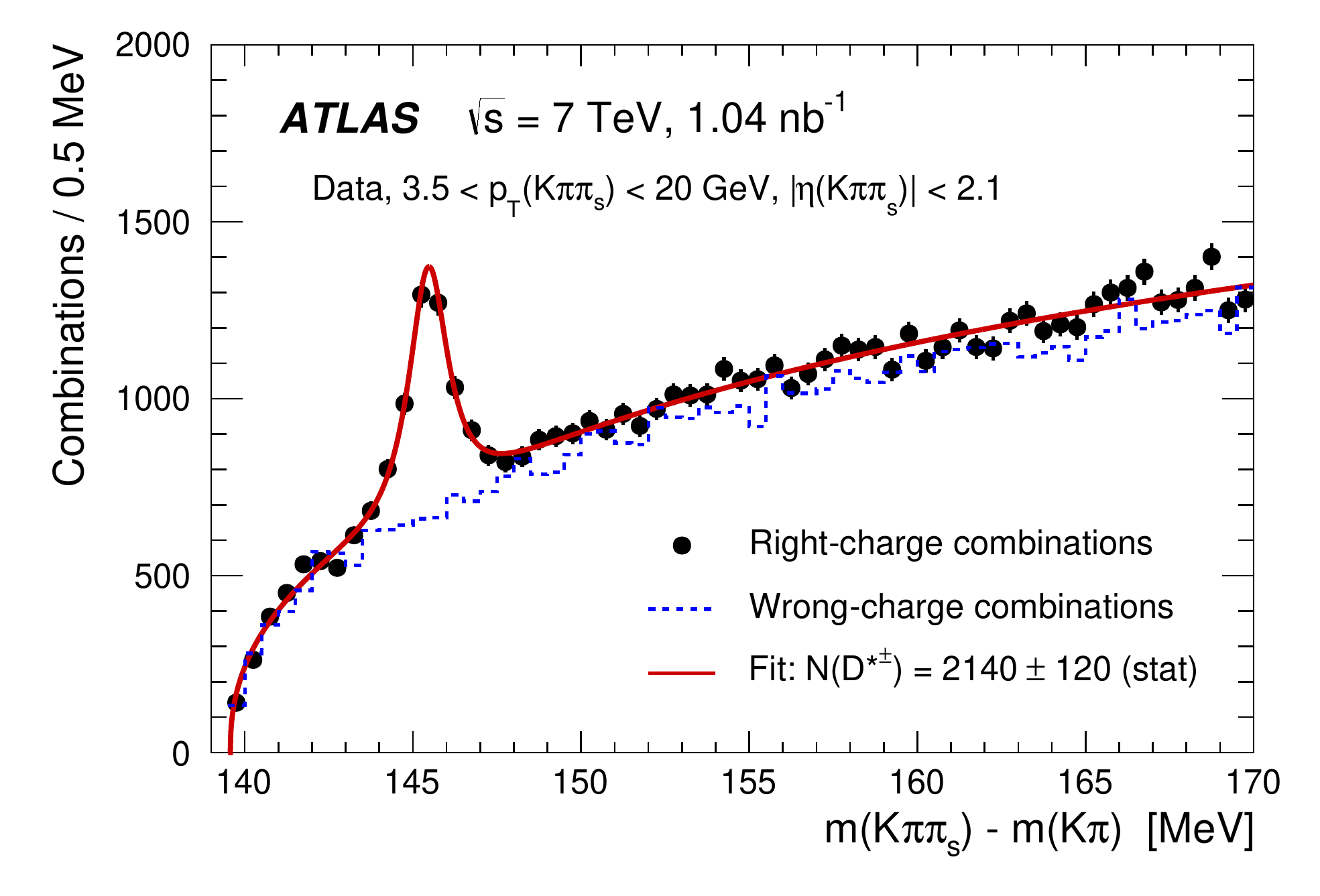}   
  \includegraphics[width=0.95\textwidth]{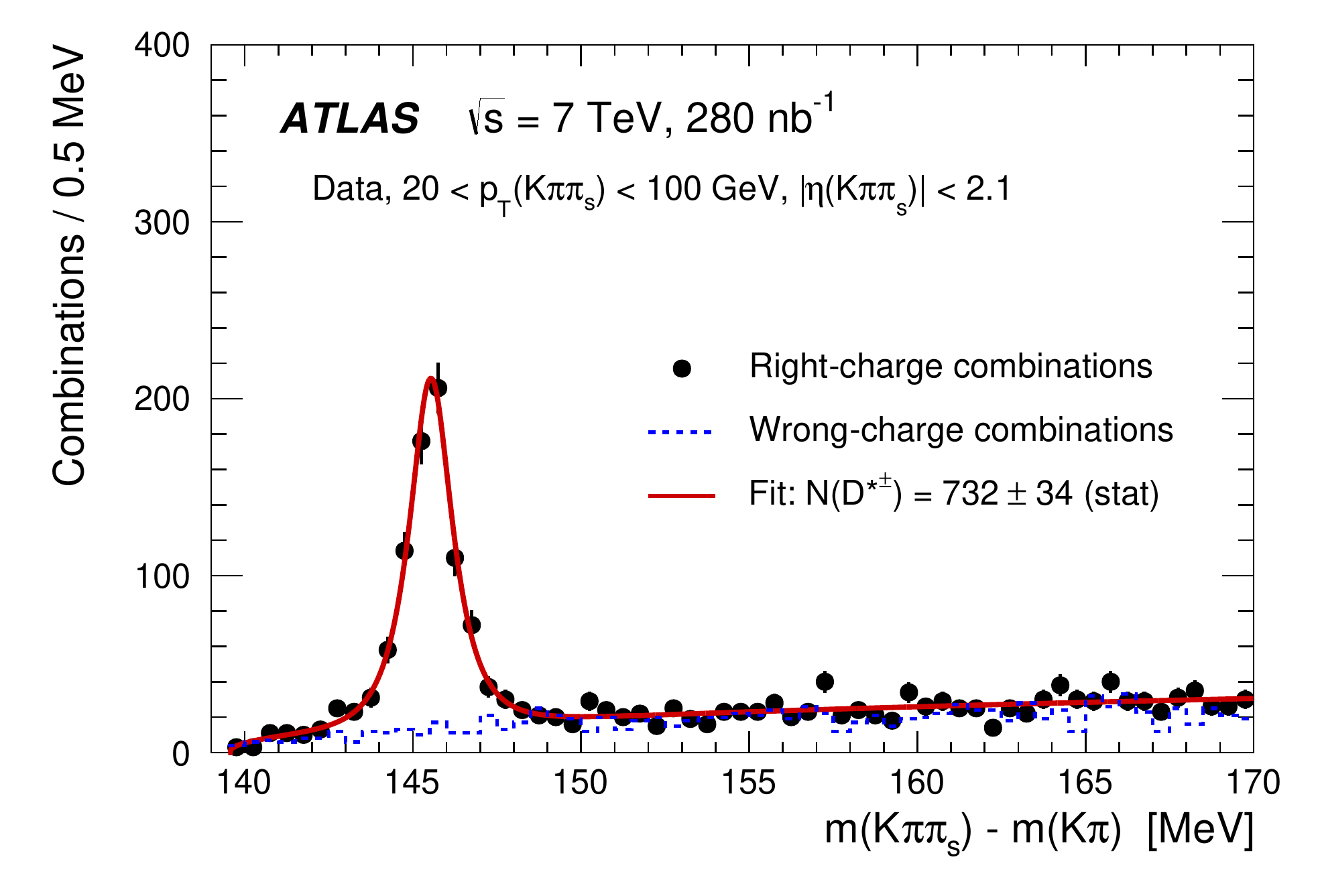}
  \caption{
The distribution of the mass difference,
$\Delta m=m(K \pi \pi_s)-m(K \pi)$,
for $\dspm$ candidates with
$3.5<\pt(\dspm)<20\gev$ (top)
and $20<\pt(\dspm)<100\gev$ (bottom).
The data are represented by the points with error bars
(statistical only).
The dashed histograms
show the distributions for wrong-charge combinations.
The solid curves represent fit results (see text).
\label{fig:ds}}
\end{figure}

Figure~\ref{fig:ds} shows
the $\Delta m$ distributions
for low-$\pt$ and high-$\pt$ $\dsp$ candidates
with $m(K \pi)$ values consistent with the
world average $D^0$ mass~\cite{pdg2014}.
To take the mass resolution into account,
the selection requirement is varied from
$1.83<m(K \pi)<1.90\gev$ for
the $\dsp$ candidates with small $|\eta|$ and $\pt$ values
to $1.78<m(K \pi)<1.95\gev$
for the $\dsp$ candidates with large $|\eta|$ and $\pt$ values.
Sizeable signals are seen around the world average value of
$m(\dsp)-m(\dz)=145.4527\pm0.0017\,$MeV~\cite{pdg2014}.
The dashed histograms show
the distributions for wrong-charge combinations,
in which both particles forming the $\dz$ candidate have the same charge
and the third particle has the opposite charge.
These distributions,
which
are quite similar to the distributions for right-charge combinations
outside of the signal region,
demonstrate the shapes of the combinatorial background components.
The $\Delta m$ distributions for the right-charge combinations
outside of the signal region
are slightly above those for the wrong-charge combinations
due to contributions from neutral-meson decays
to two particles with opposite charges, in particular due to the contribution
from $D^0$ mesons not originating from $D^{*+}\rightarrow D^0\pi^+$ decays.

The $\Delta m$ distributions are fitted
to the sum of a modified
Gaussian function~\cite{ZEUSJpsi}
describing the signal
and a threshold function
describing the non-resonant background.
The modified Gaussian function is defined as
$${\rm Gauss}^{\rm mod}\propto \exp [-0.5 \cdot x^{1+1/(1+0.5 \cdot x)}]\,,$$
where $x=|(\Delta m-m_0)/\sigma|$.
This functional form,
introduced to take into account the non-Gaussian tails of resonant signals,
describes both the data and MC signals well.
The signal position, $m_0$,
and width, $\sigma$, as well as the number of $\dsp$ mesons
are free parameters of the fit.
The threshold function has the form
$A\cdot (\Delta m -m_{\pi^+})^B\cdot\exp[C\cdot(\Delta m - m_{\pi^+})
+D\cdot(\Delta m - m_{\pi^+})^2]$,
where $m_{\pi^+}$ is the pion
mass 
and $A$, $B$, $C$ and $D$ are free parameters.
The fitted $\dspm$ yields are
$N(\dspm)=2140\pm120\,$(stat) and $N(\dspm)=732\pm34\,$(stat)
for the low-$\pt$ and high-$\pt$ ranges,
respectively.
Small admixtures ($<1\%$) to the reconstructed signals
from the $D^{*+}\rightarrow D^0\pi^+$ decays with $D^0$ decays
to final states other than $K^-\pi^+$
are taken into account in the acceptance correction procedure (Section~\ref{sec-syst}).
The combined value of the fitted
mass differences
is $145.47\pm0.03\,$(stat)$\,$MeV,
in agreement with the world average.
The widths of the signals are
$\sim$$0.6\mev$,
in agreement with the MC expectations.

\subsection{Reconstruction of $\dc$ mesons}
\label{sec-recdc}

The $\dc$ mesons are reconstructed
from the decay
$D^+ \rightarrow K^-\pi^+\pi^+$.
In each event, two tracks from same-charge particles each with
$\pt>0.8\gev$ are combined with a track
from the opposite-charge particle with $\pt>1\gev$
to form a $\dc$ candidate.
At least one of the two particles with the same charge is required
to have $\pt>1\gev$.
Only three-track combinations successfully fitted
to a common vertex are kept.
The pion mass is assigned to each of the two tracks from same-charge particles
and the kaon mass is assigned to the third track,
after which the candidate invariant mass, $m(K\pi\pi)$,
is calculated using
the track momenta refitted to the common vertex.
To suppress combinatorial background
the following requirements are used:
\begin{itemize}
\item{
$\chi^2<12$,
where $\chi^2$ is the $\dc$ candidate vertex fit quality.
}
\item{
$|d_0(\dc)|<0.15\,$mm.
}
\item{
$|z_0(\dc)\sin\theta(\dc)|<0.3\,$mm.
}
\item{
$L_{xy}(\dc)>1.2\,$mm.
The large value of the requirement on $L_{xy}(\dc)$
is motivated by the relatively large lifetime
of the $\dc$ meson~\cite{pdg2014}
and the large combinatorial background.
}
\item{
$\cos\theta^*(K)>-0.8$,
where $\theta^*(K)$
is the angle between the kaon in the $K\pi\pi$ rest frame and the $K\pi\pi$
line of flight in the laboratory frame.
}
\item{
$\cos\theta^*(\pi)>-0.85$,
where $\theta^*(\pi)$
is the angle between the pion in the $K\pi\pi$ rest frame and the $K\pi\pi$
line of flight in the laboratory frame.
}
\end{itemize}

To suppress background from $D^{*+}$ decays, combinations with
$m(K\pi\pi)-m(K\pi)<153\mev$ are removed.
The background from
$\dssp \rightarrow \phi\pi^+$, with $\phi \rightarrow K^+K^-$,
is suppressed by
rejecting any three-track $\dc$ candidate comprised
of a pair of tracks of oppositely charged particles which,
when assuming the kaon mass for both tracks,
has a two-track invariant mass
within $\pm8\mev$ of the world average $\phi$ mass~\cite{pdg2014}.
MC studies indicate that the
suppression of the $D^{*+}\rightarrow D^0\pi^+$ decays
has a negligible effect on the $\dc$ signal,
and the suppression of the $\dssp \rightarrow \phi\pi^+$ decays
rejects less than $2\%$ of the signal.
The remaining small background from $\dssp \rightarrow K^+K^-\pi^+$
decays is subtracted using the simulated reflection shape
normalised to the measured $\dssp$ rate (Section~\ref{sec-recdss}).
Smaller contributions,
affecting mass ranges outside the expected $\dc$ signal,
from the decays
$D^+_s\rightarrow \pi^+\pi^-\pi^+$,
$D^+\rightarrow K^+K^-\pi^+$,
$D^+\rightarrow \pi^+\pi^-\pi^+$ and
$D^+\rightarrow \pi^+\pi^-\pi^+\pi^0$
are subtracted
using the simulated reflection shapes
normalised to the measured $\dc$ and $\dssp$ rates.

\begin{figure}[]
  \includegraphics[width=0.95\textwidth]{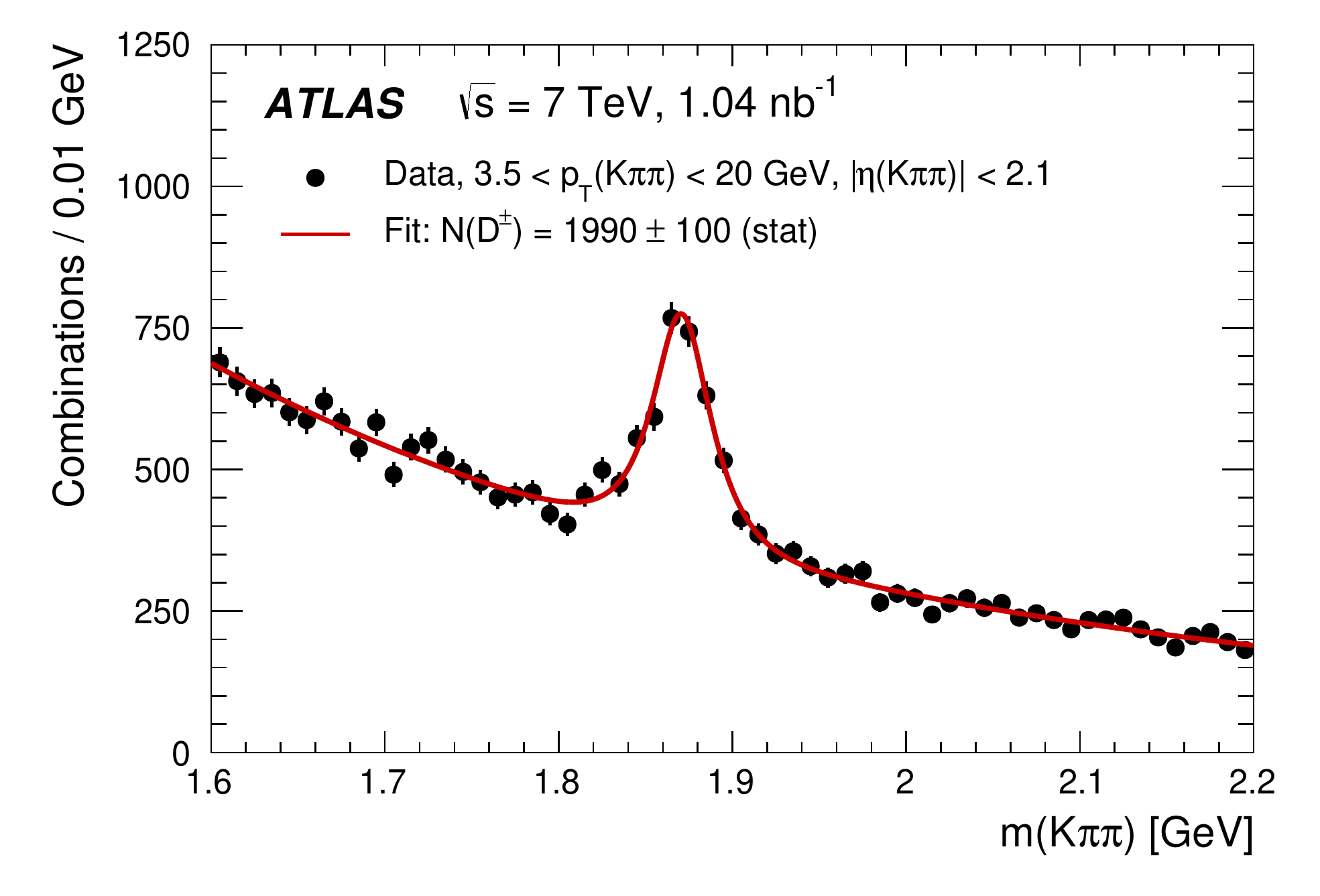}   
  \includegraphics[width=0.95\textwidth]{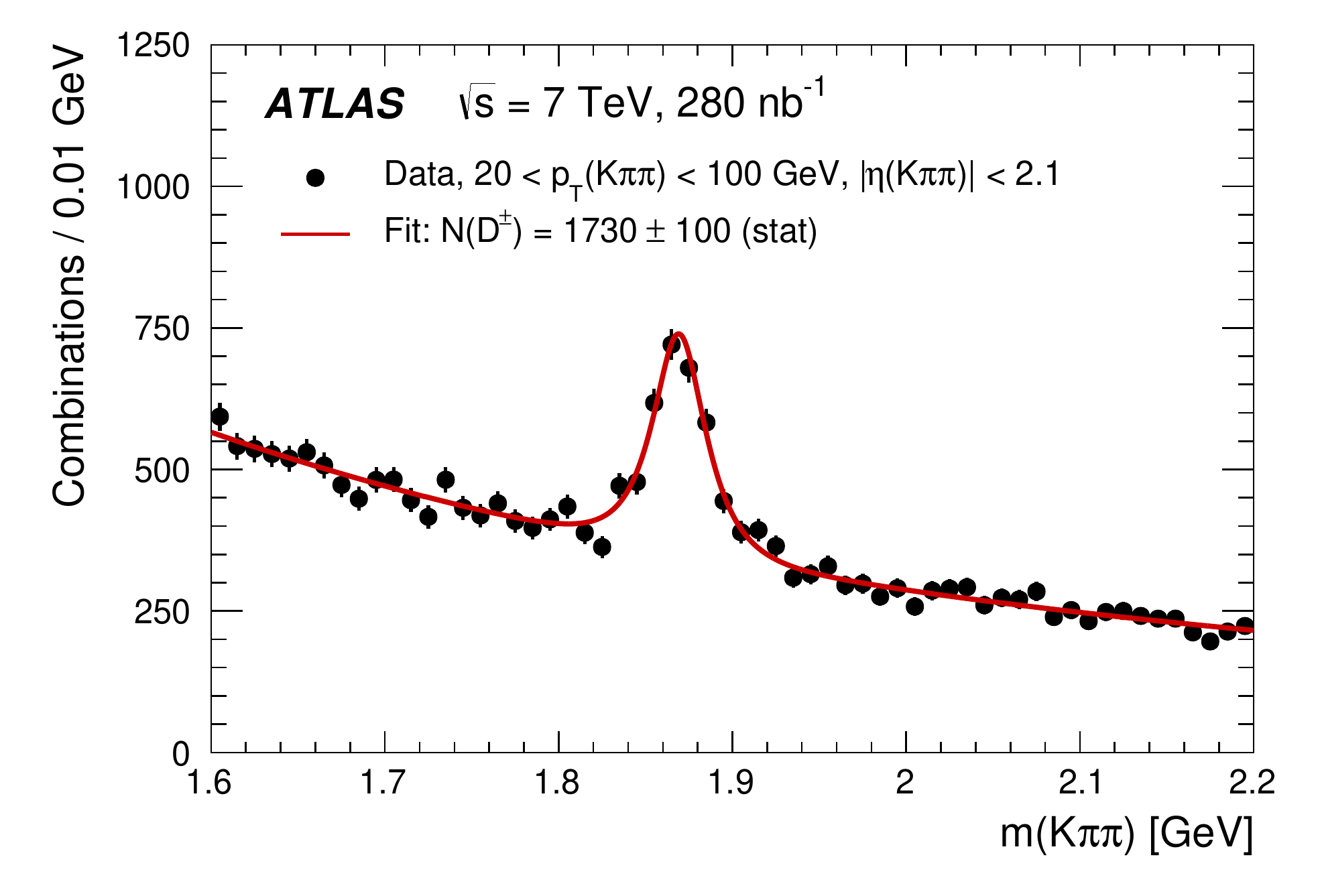}
  \caption{
The $m(K\pi\pi)$ distributions
for $\dcpm$ candidates with
$3.5<\pt(\dcpm)<20\gev$ (top)
and $20<\pt(\dcpm)<100\gev$ (bottom).
The data are represented by the points with error bars
(statistical only).
The solid curves represent fit results (see text).
\label{fig:dc}}
\end{figure}

Figure~\ref{fig:dc} shows the $m(K\pi\pi)$ distributions
for low-$\pt$ and high-$\pt$ $\dc$ candidates
after all requirements.
Sizeable signals are seen around the world average value of the $\dc$ mass,
$1869.61\pm0.10\,$MeV~\cite{pdg2014}.
The mass distributions are fitted to the sum of a modified
Gaussian function
describing the signal
and a quadratic exponential function describing the non-resonant background.
The quadratic exponential function has the form
$A\cdot \exp(B\cdot m + C\cdot m^2)$,
where
$A$, $B$ and $C$ are free parameters.
The fitted $\dcpm$ yields are
$N(\dcpm)=1990\pm100\,$(stat) and $N(\dcpm)=1730\pm100\,$(stat)
for the low-$\pt$ and high-$\pt$ ranges, respectively.
The combined $\dc$ mass value
is $1870.0\pm0.7\,$(stat)$\mev$,
in agreement with the world average.
The widths of the signals are
$\sim$$15\mev$,
in agreement with the MC expectations.

\subsection{Reconstruction of $\dssp$ mesons}
\label{sec-recdss}

The $\dssp$ mesons are reconstructed
from the decay
$\dssp \rightarrow \phi\pi^+$ with $\phi \rightarrow K^+K^-$.
In each event, tracks from particles with opposite charges and
$\pt>1\gev$
are assigned the kaon mass and
combined in pairs to form $\phi$ candidates.
Any additional track  with $\pt>1\gev$
is assigned the pion mass and combined with the $\phi$ candidate
to form a $\dssp$ candidate.
Only three-track combinations successfully fitted
to a common vertex are kept.
The $\phi$ candidate invariant mass, $m(KK)$,
and the $\dssp$ candidate invariant mass, $m(KK\pi)$,
are calculated using
the track momenta refitted to the common vertex.
To suppress combinatorial background
the following requirements are used:
\begin{itemize}
\item{
$\chi^2<12$,
where $\chi^2$ is the $\dssp$ candidate vertex fit quality.
}
\item{
$|d_0(\dssp)|<0.15\,$mm.
}
\item{
$|z_0(\dssp)\sin\theta(\dssp)|<0.3\,$mm.
}
\item{
$L_{xy}(\dssp)>0.4\,$mm.
}
\item{
$-0.8 < \cos\theta^*(\pi)<0.7$, where $\theta^*(\pi)$
is the angle between the pion in the $K K\pi$ rest frame and the $K K\pi$
line of flight in the laboratory frame.
}
\item{
$|\cos^3\theta^{\prime}(K)|>0.2$, where $\theta^{\prime}(K)$
is the angle between either of the kaons and the pion in
the $KK$ rest frame.
The decay of the pseudoscalar $\dssp$ meson to the $\phi$ (vector)
plus $\pi^+$ (pseudoscalar) final state results in an alignment of
the spin of the $\phi$ meson transverse to the direction of
motion of the $\phi$ relative to the $\dssp$.
Consequently, the distribution of $\cos\theta^{\prime}(K)$
follows a $\cos^2\theta^{\prime}(K)$ shape, implying a uniform
distribution for $\cos^3\theta^{\prime}(K)$.
In contrast, the $\cos\theta^{\prime}(K)$ distribution of
the combinatorial background is uniform and its $\cos^3\theta^{\prime}(K)$
distribution peaks at zero. The requirement suppresses the background significantly
while reducing the signal by $20\%$.
}
\end{itemize}

Small contributions,
affecting mass ranges outside the expected $\dssp$ signal,
from the decays
$D^+_s\rightarrow \phi K^+$,
$D^+_s\rightarrow \phi \pi^+\pi^0$,
$D^+\rightarrow \phi \pi^+\pi^0$ and
$D^+\rightarrow K^-\pi^+\pi^+$
are subtracted using the simulated reflection shapes
normalised to the measured $\dc$ and $\dssp$ rates.

\begin{figure}[]
  \includegraphics[width=0.95\textwidth]{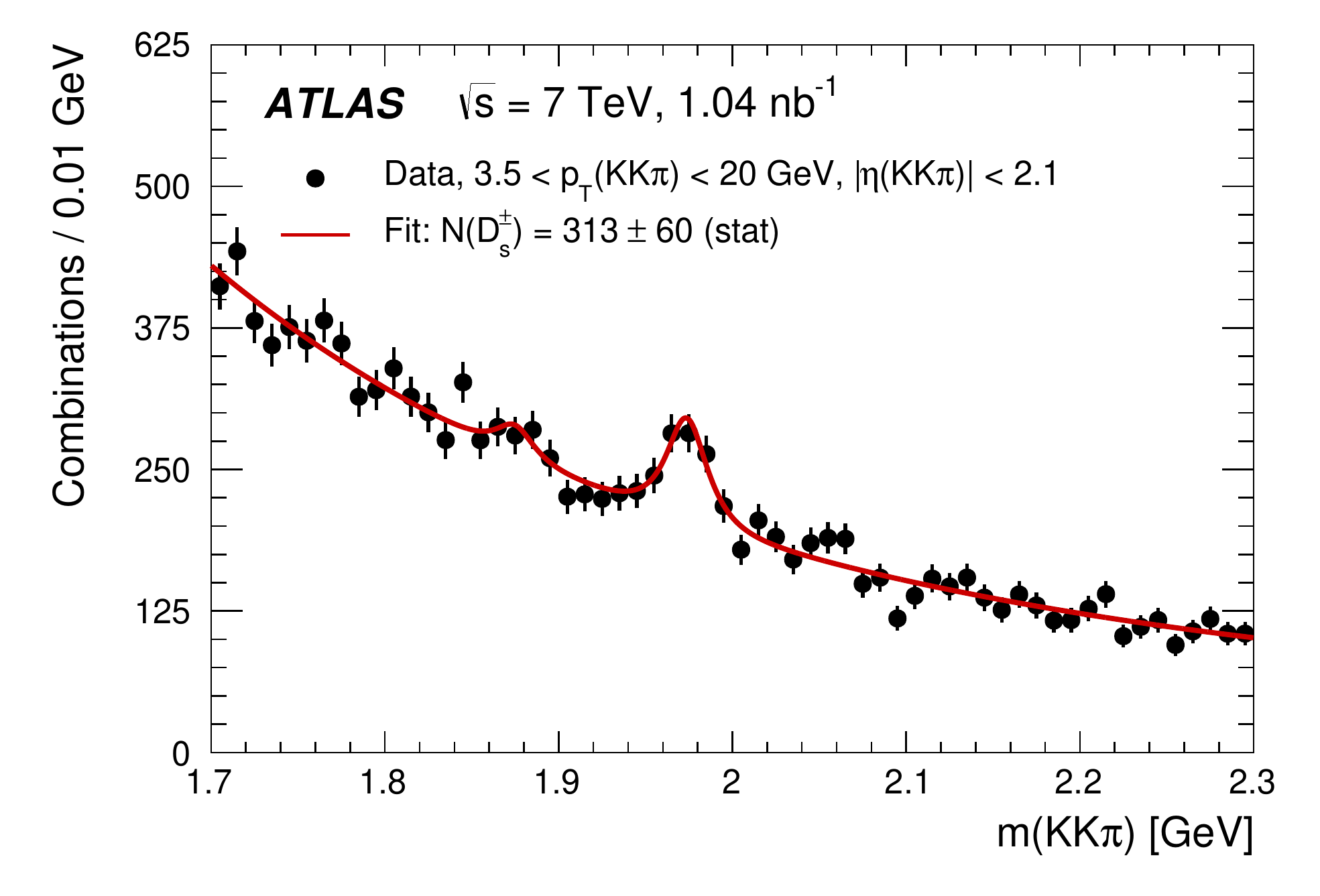}   
  \includegraphics[width=0.95\textwidth]{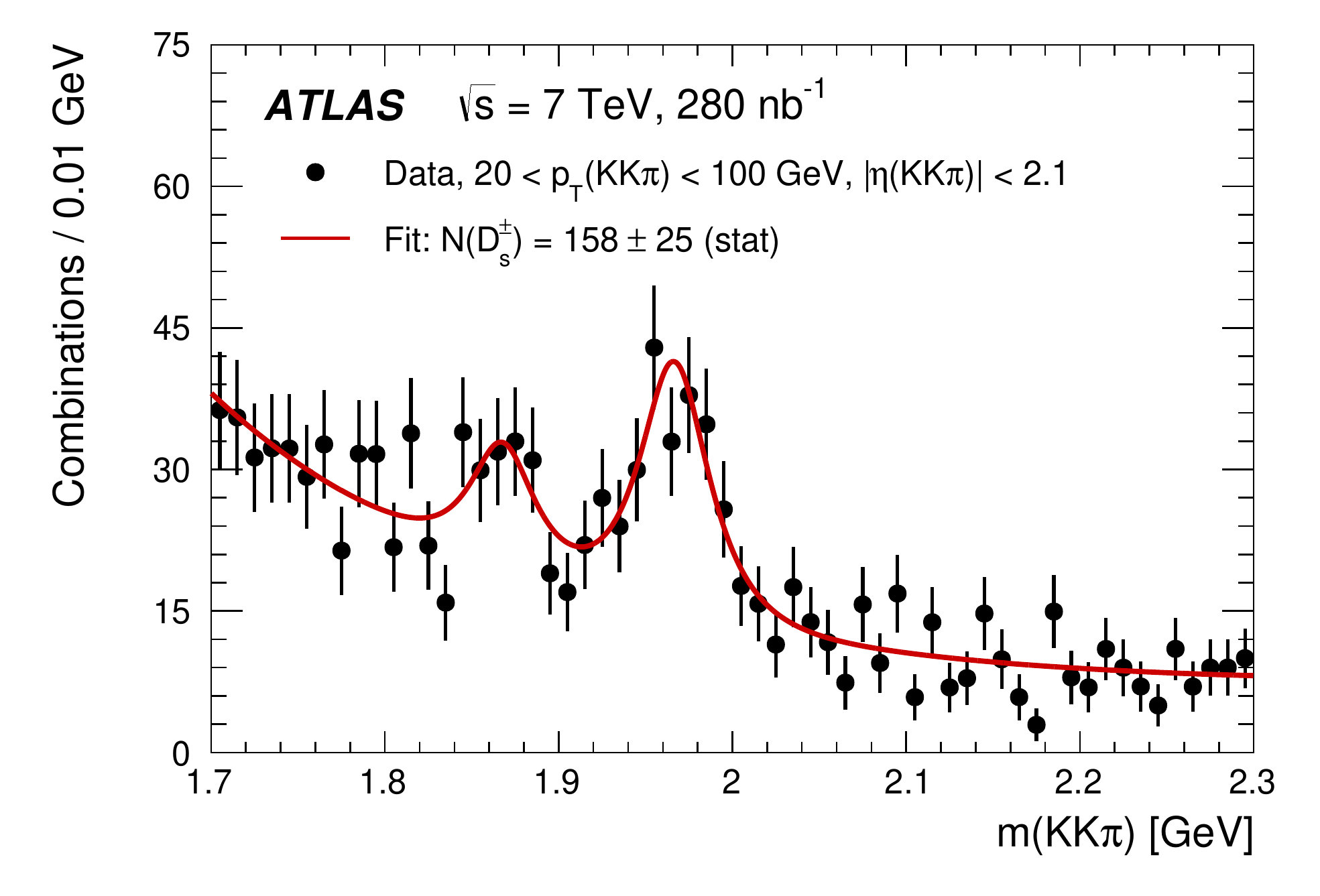}
  \caption{
The $m(K K\pi)$ distributions
for $\dsspm$ candidates with
$3.5<\pt(\dsspm)<20\gev$ (top)
and $20<\pt(\dsspm)<100\gev$ (bottom).
Small signals visible around the world average value of $m(\dc)$
are
from the decay
$\dc \rightarrow \phi\pi^+$ with $\phi \rightarrow K^+K^-$.
The data are represented by the points with error bars
(statistical only).
The solid curves represent the fit results (see text).
\label{fig:dss}}
\end{figure}

Figure~\ref{fig:dss} shows the $m(K K\pi)$ distributions
for low-$\pt$ and high-$\pt$ $\dssp$ candidates
with $m(KK)$
within $\pm7\mev$ of the world average $\phi$ mass~\cite{pdg2014}.
Sizeable signals are seen around the world average value of the $\dssp$ mass,
$1968.30\pm0.11\,$MeV~\cite{pdg2014}.
Smaller signals are visible around the world average value of $m(\dc)$,
as expected
from the decay
$\dc \rightarrow \phi\pi^+$ with $\phi \rightarrow K^+K^-$.

The $m(KK\pi)$ distributions are fitted to the sum
of two modified
Gaussian functions
describing the $\dssp$ and $\dc$ signals
and a quadratic exponential function describing the non-resonant background.
For the small $\dc$ signals,
the signal positions are fixed
to the $\dssp$ signal positions minus the world average value
of $m(\dssp)-m(\dc)$~\cite{pdg2014}, and
their widths are fixed using the $\dssp$ signal widths
and the MC ratio of the  $\dc$ and $\dssp$ widths.
The fitted $\dsspm$ yields are
$N(\dsspm)=313\pm60\,$(stat) and $N(\dsspm)=158\pm25\,$(stat)
for the low-$\pt$ and high-$\pt$ ranges, respectively.
The combined $\dssp$ mass value
is $1971.2\pm2.0\,$(stat)$\,$MeV,
in agreement with the world average.
The widths of the signals are
$\sim$$15\mev$,
in agreement with the MC expectations.

\section{\label{secCorrections}Data correction and systematic uncertainties}
\label{sec-syst}

The visible $D$ production cross sections are
measured
for the process  $p p\rightarrow D X$
in the kinematic region
$3.5<\pt(D)<100\gev$ and $|\eta(D)|<2.1$.
The cross section for a given charmed meson is calculated
in the low-$\pt$ range, $3.5<\pt(D)<20\gev$,
and high-$\pt$ range, $20<\pt(D)<100\gev$,
from

\begin{equation}
\sigma_{p p\rightarrow D X}=\frac{N(D)}{\acc \cdot \lum \cdot \bran}\,,
\end{equation}
where $N(D)$ is the number of reconstructed charmed mesons
with positive and negative charges,
$\acc$ is the reconstruction acceptance obtained from the MC sample,
$\lum$ is the integrated luminosity
and $\bran$ is the branching fraction or the product of the
branching fractions for the decay channel
used in the reconstruction.
The reconstruction acceptance takes into account
efficiencies, migrations and
small remaining admixtures in the reconstructed signals from
other decay modes.
To calculate the $\dsp$ and $\dc$ production cross sections,
the world average $\bran$ values~\cite{pdg2014} are used.
For $\dssp$, the measurement by the CLEO experiment~\cite{cleodss}
of the partial $\dssp\ra K^+K^-\pi^+$
branching fractions, with a kaon-pair mass within various intervals
around the world average $\phi$ meson mass, is used.
Interpolating between the partial branching fractions,
measured for the $\pm5\,$MeV and $\pm10\,$MeV intervals,
yields the value $(1.85\pm0.11)\%$ for the $\pm7\,$MeV interval used in this analysis.

The differential cross sections $\dr\sigma/\dr\pt$ and $\dr\sigma/\dr |\eta|$
are calculated for $\dsp$ and $\dc$ production\footnote
{For $\dssp$ production, the differential cross sections
are not calculated due to insufficient sample size.}
in nine bins in $\pt$ $(3.5-5;~5-6.5;~6.5-8;~8-12;~12-20;~20-30;~30-40;~40-60;~60-100\,\rm{GeV})$,
and five bins in $|\eta|$ $(0-0.2;~0.2-0.5;~0.5-0.8;~0.8-1.3;~1.3-2.1)$
for both the low-$\pt$ and high-$\pt$ ranges.
To obtain the differential cross section in a given bin, the visible
cross section in the bin is divided by the bin width.
The numbers of $\dsp$ and $\dc$ mesons in each bin are obtained using
the same procedure as that
described in Section~\ref{secReconstruction}.

The following groups of systematic uncertainty sources
are considered:
\begin{itemize}
\item{$\{\delta_1\}$
The uncertainty of the jet trigger efficiencies.
It
is estimated using data--MC comparisons with independent trigger selections.
}
\item{$\{\delta_2\}$
The uncertainty of the track reconstruction and selection~\cite{ATLAStracks}.
It is dominated by the uncertainty
on the description
of the detector material
in the MC simulation.
The uncertainty is calculated taking into account
the $\pt$ and $\eta$ distributions of the $D$ decay products.
}
\item{$\{\delta_3\}$
The uncertainty of the $D$ meson selection efficiency.
It is determined by varying
the MC reconstruction resolutions
for the variables
used in the selection of
the $D$ meson
by amounts reflecting possible
differences between the data and MC.
For the $\pt(D^{*+}, D^+, D_s^+)/ \sum \pt(\rm{track}) > 0.05$
requirement,
the uncertainty is determined by repeating all calculations
without this requirement.
}
\item{$\{\delta_4\}$
The uncertainty related to the $D$ signal extraction procedures.
It is determined by varying the background parameterisations
and the ranges used for the signal fits.
In addition, in the $\dc$ signal extraction procedure, the normalisation of
the subtracted
$\dssp\rightarrow K^-K^+\pi^+$ reflection is varied
in the combined range of the normalisation statistical uncertainty
and normalisation uncertainty propagated
from the branching fraction uncertainties~\cite{pdg2014}.
In the $\dssp$ signal extraction procedure,
the constraints used for the
small $\dc$ signals are varied
in the ranges of the MC statistical uncertainty
for the ratio of the $\dc$ and $\dssp$ widths
and the uncertainty of world average value
of $m(\dssp)-m(\dc)$~\cite{pdg2014}.
}
\item{$\{\delta_5\}$
The model dependence of the acceptance corrections.
It is obtained by varying in the MC simulation:
\begin{itemize}
\item{}the $\pt(D)$ and
$|\eta(D)|$ distributions
while
preserving agreement with the data distributions,
\item{}the
relative beauty contribution
in the range allowed by the
$b$-hadron cross-section measurement~\cite{ATLASDsmuon},
\item{}the lifetimes of charmed
($D^+$, $D^0$, $D_s^+$)
and beauty
($B^+$, $B^0$, $B^0_s$, $\Lambda_b^0$)
hadrons
in the ranges of their uncertainties~\cite{pdg2014}.
\end{itemize}
}
\item{$\{\delta_6\}$
The uncertainty
of the acceptance corrections
related to the MC statistical uncertainty.
}
\item{$\{\delta_7\}$
The uncertainty of the luminosity measurement~\cite{ATLASlumi}.
}
\item{$\{\delta_8\}$
The uncertainty of
the branching fractions~\cite{pdg2014,cleodss} used in Eq.~(1).
}
\end{itemize}

\begin{table}[hbt!]
\begin{center}
\renewcommand{\arraystretch}{1.4}
\begin{tabular}{c|c|c|c|c|c|c} \hline
Source & \multicolumn{2}{c|}{$\sv(D^{*\pm}$)}
       & \multicolumn{2}{c|}{$\sv(D^{\pm}$)}
       & \multicolumn{2}{c}{$\sv(D_s^{\pm}$)} \\
\hline
 & Low-$p_{\rm T}$ & High-$p_{\rm T}$ & Low-$p_{\rm T}$ & High-$p_{\rm T}$ & Low-$p_{\rm T}$ & High-$p_{\rm T}$ \\
\hline
\hfill Trigger ($\delta_1$) & - & $\,^{+0.9}_{-1.0}\%$& - & $\,^{+0.9}_{-1.0}\%$& - & $\,^{+0.9}_{-1.0}\%$ \\
\hfill Tracking ($\delta_2$) & $\pm7.8\%$ & $\pm7.4\%$ & $\pm7.7\%$ & $\pm7.4\%$ & $\pm7.6\%$ & $\pm7.4\%$ \\
\hfill $D$ selection ($\delta_3$) & $\,^{+2.8}_{-1.6}\%$ & $\,^{+1.7}_{-1.4}\%$ & $\,^{+1.6}_{-1.0}\%$ & $\,^{+0.9}_{-0.6}\%$ & $\,^{+2.6}_{-1.6}\%$ & $\,^{+1.1}_{-0.9}\%$ \\
\hfill Signal fit ($\delta_4$) & $\pm1.3\%$ & $\pm0.9\%$ & $\pm1.3\%$ & $\pm1.5\%$ & $\pm6.4\%$ & $\pm5.3\%$ \\
\hfill Modelling ($\delta_5$) & $\,^{+1.0}_{-1.7}\%$ & $\,^{+2.7}_{-2.3}\%$ & $\,^{+2.3}_{-2.6}\%$ & $\,^{+2.9}_{-2.4}\%$ & $\,^{+1.7}_{-2.4}\%$ & $\,^{+2.8}_{-2.4}\%$ \\
\hfill Size of MC sample ($\delta_6$) & $\pm0.6\%$ & $\pm0.9\%$ & $\pm0.8\%$ & $\pm0.8\%$ & $\pm2.9\%$ & $\pm3.1\%$ \\
\hfill Luminosity ($\delta_7$) & $\pm3.5\%$ & $\pm3.5\%$ & $\pm3.5\%$ & $\pm3.5\%$ & $\pm3.5\%$ & $\pm3.5\%$ \\
\hfill Branching fraction ($\delta_8$) & $\pm1.5\%$ & $\pm1.5\%$ & $\pm2.1\%$ & $\pm2.1\%$ & $\pm5.9\%$ & $\pm5.9\%$ \\
\hline
\end{tabular}
\caption{
Systematic uncertainties for measurements of visible
low-$\pt$, $3.5<\pt(D)<20\gev$,
and high-$\pt$, $20<\pt(D)<100\gev$,
cross sections of $D^{*\pm}$, $D^\pm$ and $D^\pm_s$
production with $|\eta|<2.1$.
}
\label{tab:syst}
\end{center}
\end{table}

The systematic uncertainties are summarised in Table~\ref{tab:syst}.
Contributions from
the systematic uncertainties $\delta_1-\delta_6$, calculated
for visible cross sections and all bins of the differential cross sections,
are added
in quadrature separately for positive and negative variations.
Uncertainties linked with the luminosity measurement
($\delta_7$) and
branching fractions ($\delta_8$) are quoted separately
for the measured visible cross sections.
For differential cross sections,
the $\delta_7$ and $\delta_8$
uncertainties are not included
in Tables~\ref{tab:xsdpt}--\ref{tab:xsdeta2}
and Figs.~\ref{fig:sig_dpt}--\ref{fig:sig_detahpt}.

\section{\label{secCrosssections}Production cross sections of charmed mesons}
\label{sec-xsec}

The visible cross sections of $D$ meson production
in $pp$ collisions at $\sqrt{s}=7\,$TeV
for $|\eta(D)|<2.1$
in the low-$\pt$ range, $3.5<\pt(D)<20\gev$,
are measured to be
$$\sv(D^{*\pm})   =331\pm18\,({\rm stat})\pm28\,({\rm syst})\pm12\,({\rm lum})\pm\,\,5\,({\rm br})\,\mub\,,$$
$$\sv(D^{\pm})\,\,=328\pm16\,({\rm stat})\pm27\,({\rm syst})\pm11\,({\rm lum})\pm\,\,7\,({\rm br})\,\mub\,,$$
$$\sv(D_s^{\pm})\,\,=160\pm31\,({\rm stat})\pm17\,({\rm syst})\pm\,\,\,6\,({\rm lum})\pm10\,({\rm br})\,\mub\,,$$
where the last two uncertainties are due to
those on the luminosity measurement and
the charmed meson decay branching fractions.

The POWHEG+PYTHIA predictions are
$$\sv(D^{*\pm})=158^{+176}_{\,\,\,-81}\,({\rm scale})^{+15}_{-16}\,(m_Q)\,^{+14}_{-13}\,({\rm PDF}\oplus\alpha_s)^{+19}_{-16}\,({\rm hadr})\,\mub\,,$$
$$\sv(D^{\pm})\,\,=134^{+145}_{\,\,\,-67}\,({\rm scale})^{+12}_{-13}\,(m_Q)\,^{+12}_{-11}\,({\rm PDF}\oplus\alpha_s)^{+21}_{-12}\,({\rm hadr})\,\mub\,,$$
$$\sv(D_s^{\pm})\,=\,\,\,\,62^{\,+63\,}_{\,-29\,}\,({\rm scale})\pm6\,(m_Q)\pm5\,({\rm PDF}\oplus\alpha_s)^{\,+7\,}_{\,-8\,}\,({\rm hadr})\,\mub\,,$$
where the last uncertainty is due to that on hadronisation
(see Section~\ref{secNLO}).
The FONLL predictions for $\dsp$ and $\dc$ are
$$\sv(D^{*\pm})=202^{+119}_{\,\,\,-73}\,({\rm scale})^{+36}_{-27}\,(m_Q)\pm21\,({\rm PDF})\pm5\,({\rm ff})\,\mub\,,$$
$$\sv(D^{\pm})\,\,=174^{\,\,\,+99}_{\,\,\,-60}\,({\rm scale})^{+33}_{-24}\,(m_Q)\pm18\,({\rm PDF})\pm7\,({\rm ff})\,\mub\,,$$
where the last uncertainty is due to that on the fragmentation function.
The FONLL predictions for $\dssp$ production are currently not available.

\begin{table}[hbt!]
\begin{center}
\renewcommand{\arraystretch}{1.4}
\begin{tabular}{c|c|c|c|c|c|c} \hline
 & \multicolumn{2}{c|}{$\sv(D^{*\pm}$)}
 & \multicolumn{2}{c|}{$\sv(D^{\pm}$)}
 & \multicolumn{2}{c}{$\sv(D_s^{\pm}$)} \\
\hline
Range & low-$p_{\rm T}$ & high-$p_{\rm T}$ & low-$p_{\rm T}$ & high-$p_{\rm T}$ & low-$p_{\rm T}$ & high-$p_{\rm T}$ \\
$[$units$]$ & $[\mub ]$ & [nb] & $[\mub ]$ & [nb] & $[\mub ]$ & [nb] \\
\hline
ATLAS & $331\pm 36$ & $988\pm 100$ & $328\pm 34$ & $888\pm 97$ & $160\pm 37$ & $512\pm 104$ \\
\hline
GM-VFNS & $340^{+130}_{-150} $ & $1000^{+120}_{-150} $ & $350^{+150}_{-160} $ & $980^{+120}_{-150}$ & $147^{+54}_{-66} $ & $470^{+56}_{-69}$ \\
FONLL & $202^{+125}_{\,\,\,-79} $ & $753^{+123}_{-104} $ & $174^{+105}_{\,\,\,-66} $ & $617^{+103}_{\,\,\,-86}$ & - & - \\
POWHEG+PYTHIA & $158^{+179}_{\,\,\,-85} $ & $600^{+300}_{-180} $ & $134^{+148}_{\,\,\,-70} $ & $480^{+240}_{-130}$ & $62^{+64}_{-31} $ & $225^{+114}_{\,\,\,-69}$ \\
POWHEG+HERWIG & $137^{+147}_{\,\,\,-72} $ & $690^{+380}_{-160} $ & $121^{+129}_{\,\,\,-64} $ & $580^{+280}_{-140}$ & $51^{+50}_{-25} $ & $268^{+107}_{\,\,\,-62}$ \\
MC@NLO & $157^{+125}_{\,\,\,-72} $ & $980^{+460}_{-290} $ & $140^{+112}_{\,\,\,-65} $ & $810^{+390}_{-260}$ & $58^{+42}_{-25} $ & $345^{+175}_{\,\,\,-87}$ \\
\hline
\end{tabular}
\caption{
The visible
low-$\pt$, $3.5<\pt(D)<20\gev$,
and high-$\pt$, $20<\pt(D)<100\gev$,
cross sections of $D^{*\pm}$, $D^\pm$ and $D^\pm_s$
production with $|\eta|<2.1$.
The measurements are compared with the
GM-VFNS~\cite{gmvfns1,gmvfns2,gmvfns3},
FONLL~\cite{fonll1,fonll2,fonll3,fonllweb},
POWHEG+PYTHIA~\cite{powhegHQ,pythia6},
POWHEG+HERWIG~\cite{powhegHQ,herwig} and
MC@NLO~\cite{mcatnloHQ,herwig} predictions.
The data uncertainties
are the total uncertainties obtained as sums in quadrature
of the statistical, systematic, luminosity and branching-fraction uncertainties.
The prediction uncertainties
are the total uncertainties obtained as sums in quadrature
of all considered sources of the theoretical uncertainty (see text).
}
\label{tab:xsvis}
\end{center}
\end{table}

The visible cross sections of $D$ meson production
in $pp$ collisions at $\sqrt{s}=7\,$TeV
for $|\eta(D)|<2.1$
in the high-$\pt$ range, $20<\pt(D)<100\gev$,
are measured to be
$$\sv(D^{*\pm})=988\pm45\,({\rm stat})\pm81\,({\rm syst})\pm35\,({\rm lum})\pm15\,({\rm br})\,{\rm nb}\,,$$
$$\sv(D^{\pm})\,\,=888\pm53\,({\rm stat})\pm73\,({\rm syst})\pm31\,({\rm lum})\pm18\,({\rm br})\,{\rm nb}\,,$$
$$\sv(D_s^{\pm})\,\,=512\pm83\,({\rm stat})\pm52\,({\rm syst})\pm18\,({\rm lum})\pm30\,({\rm br})\,{\rm nb}\,.$$
The POWHEG+PYTHIA predictions are
$$\sv(D^{*\pm})=600^{+269}_{-137}\,({\rm scale})^{+15}_{-21}\,(m_Q)^{+25}_{-34}\,({\rm PDF}\oplus\alpha_s)^{+126}_{-111}\,({\rm hadr})\,{\rm nb}\,,$$
$$\sv(D^{\pm})\,\,=480^{+208}_{-109}\,({\rm scale})^{\,\,\,+6}_{-11}\,(m_Q)^{+20}_{-27}\,({\rm PDF}\oplus\alpha_s)^{+121}_{\,\,\,-71}\,({\rm hadr})\,{\rm nb}\,,$$
$$\sv(D_s^{\pm})\,\,=225^{+106}_{\,\,\,-47}\,({\rm scale})^{\,\,+9\,}_{\,\,-8\,}\,(m_Q)^{\,\,\,+9}_{-13}\,({\rm PDF}\oplus\alpha_s)^{\,\,\,+40\,}_{\,\,\,-49\,}\,({\rm hadr})\,{\rm nb}\,.$$
The FONLL predictions for $\dsp$ and $\dc$ are
$$\sv(D^{*\pm})=753^{+116}_{\,\,\,-98}\,({\rm scale})^{+28}_{-18}\,(m_Q)\pm41\,({\rm PDF})\pm17\,({\rm ff})\,\mub\,,$$
$$\sv(D^{\pm})\,\,=617^{\,\,\,+92\,}_{\,\,\,-78\,}\,({\rm scale})^{+37}_{-21}\,(m_Q)\pm33\,({\rm PDF})\pm23\,({\rm ff})\,\mub\,.$$

\begin{table}[hbt!]
\begin{center}
\renewcommand{\arraystretch}{1.4}
\begin{tabular}{c|c|c} \hline
$\pt$ range &
$\dr\sig/\dr\pt (D^{*\pm})$ [$\mub$/GeV] &
$\dr\sig/\dr\pt (D^{\pm})$ [$\mub$/GeV] \\
\hline
$3.5-5.0$ & $145\pm 15 \pm 14$ & $127\pm 13 \pm 12$ \\
$5.0-6.5$ & $43.4\pm 4.2 \pm 3.6$ & $51.9\pm 4.3 \pm 4.2$ \\
$6.5-8.0$ & $20.8\pm 1.9 \pm 1.7$ & $20.0\pm 2.3 \pm 1.6$ \\
$~~8-12$ & $6.34\pm 0.50 \pm 0.51$ & $6.29\pm 0.56 \pm 0.51$ \\
$12-20$ & $(757\pm 101 \pm 65)\times10^{-3}$ & $(583\pm 88 \pm 50)\times10^{-3}$ \\
$20-30$ & $(78.8\pm 5.6 \pm 6.4)\times10^{-3}$ & $(73.6\pm 5.5 \pm 5.9)\times10^{-3}$ \\
$30-40$ & $(13.3\pm 1.2 \pm 1.2)\times10^{-3}$ & $(11.9\pm 1.2 \pm 1.0)\times10^{-3}$ \\
$40-60$ & $(2.52\pm 0.21 \pm 0.20)\times10^{-3}$ & $(2.05\pm 0.18 \pm 0.16)\times10^{-3}$ \\
$~~60-100$ & $(131\pm 31 \pm 11)\times10^{-6}$ & $(175\pm 41 \pm 15)\times10^{-6}$ \\
\hline
\end{tabular}
\caption{
The measured differential cross sections
$\dr\sig/\dr\pt$
of $D^{*\pm}$ and $D^\pm$
production with $|\eta|<2.1$.
The first and second errors
are the statistical and systematic uncertainties, respectively.
The systematic uncertainties
corresponding to the tracking ($\delta_2$) uncertainties (Table~\ref{tab:syst})
are strongly correlated.
The fully correlated
uncertainties linked with the luminosity measurement ($3.5\%$)
and branching fractions ($1.5\%$ and $2.1\%$
for $D^{*\pm}$ and $D^\pm$, respectively)
are not shown.
}
\label{tab:xsdpt}
\end{center}
\end{table}

\begin{table}[hbt!]
\begin{center}
\renewcommand{\arraystretch}{1.4}
\begin{tabular}{c|c|c} \hline
$|\eta|$ range &
$\dr\sig/\dr|\eta| (D^{*\pm})$ [$\mub$] &
$\dr\sig/\dr|\eta| (D^{\pm})$ [$\mub$] \\
\hline
$0.0-0.2$ & $176\pm 21 \pm 14$ & $165\pm 20 \pm 13$ \\
$0.2-0.5$ & $158\pm 17 \pm 12$ & $164\pm 16 \pm 13$ \\
$0.5-0.8$ & $149\pm 15 \pm 12$ & $165\pm 15 \pm 13$ \\
$0.8-1.3$ & $156\pm 14 \pm 14$ & $157\pm 17 \pm 13$ \\
$1.3-2.1$ & $171\pm 23 \pm 19$ & $142\pm 19 \pm 18$ \\
\hline
\end{tabular}
\caption{
The measured differential cross sections
$\dr\sig/\dr|\eta|$
of $D^{*\pm}$ and $D^\pm$
production with $3.5<\pt<20\,$GeV.
The first and second errors
are the statistical and systematic uncertainties, respectively.
The systematic uncertainty fractions
corresponding to the tracking ($\delta_2$) uncertainties (Table~\ref{tab:syst})
are strongly correlated.
The fully correlated
uncertainties linked with the luminosity measurement ($3.5\%$)
and branching fractions ($1.5\%$ and $2.1\%$
for $D^{*\pm}$ and $D^\pm$, respectively)
are not shown.
}
\label{tab:xsdeta1}
\end{center}
\end{table}

\begin{table}[hbt!]
\begin{center}
\renewcommand{\arraystretch}{1.4}
\begin{tabular}{c|c|c} \hline
$|\eta|$ range &
$\dr\sig/\dr|\eta| (D^{*\pm})$ [nb] &
$\dr\sig/\dr|\eta| (D^{\pm})$ [nb] \\
\hline
$0.0-0.2$ & $591\pm 66 \pm 46$ & $579\pm 80 \pm 46$ \\
$0.2-0.5$ & $584\pm 54 \pm 46$ & $543\pm 51 \pm 42$ \\
$0.5-0.8$ & $638\pm 55 \pm 49$ & $510\pm 51 \pm 42$ \\
$0.8-1.3$ & $446\pm 43 \pm 35$ & $408\pm 46 \pm 33$ \\
$1.3-2.1$ & $358\pm 49 \pm 40$ & $350\pm 65 \pm 39$ \\
\hline
\end{tabular}
\caption{
The measured differential cross sections
$\dr\sig/\dr|\eta|$
of $D^{*\pm}$ and $D^\pm$
production with $20<\pt<100\,$GeV.
The first and second errors
are the statistical and systematic uncertainties, respectively.
The systematic uncertainty fractions
corresponding to the tracking ($\delta_2$) uncertainties (Table~\ref{tab:syst})
are strongly correlated.
The fully correlated
uncertainties linked with the luminosity measurement ($3.5\%$)
and branching fractions ($1.5\%$ and $2.1\%$
for $D^{*\pm}$ and $D^\pm$, respectively)
are not shown.
}
\label{tab:xsdeta2}
\end{center}
\end{table}

The visible
low-$\pt$
and high-$\pt$
$D^{*\pm}$, $D^\pm$ and $D^\pm_s$
production cross sections
are compared in
Table~\ref{tab:xsvis}
with the NLO QCD predictions.
The FONLL, MC@NLO and POWHEG predictions are consistent with the data
within the large theoretical uncertainties,
with the central values
of the predictions lying
below the measurements.
The GM-VFNS predictions agree with data.

\begin{figure}[]
  \includegraphics[width=0.9\textwidth]{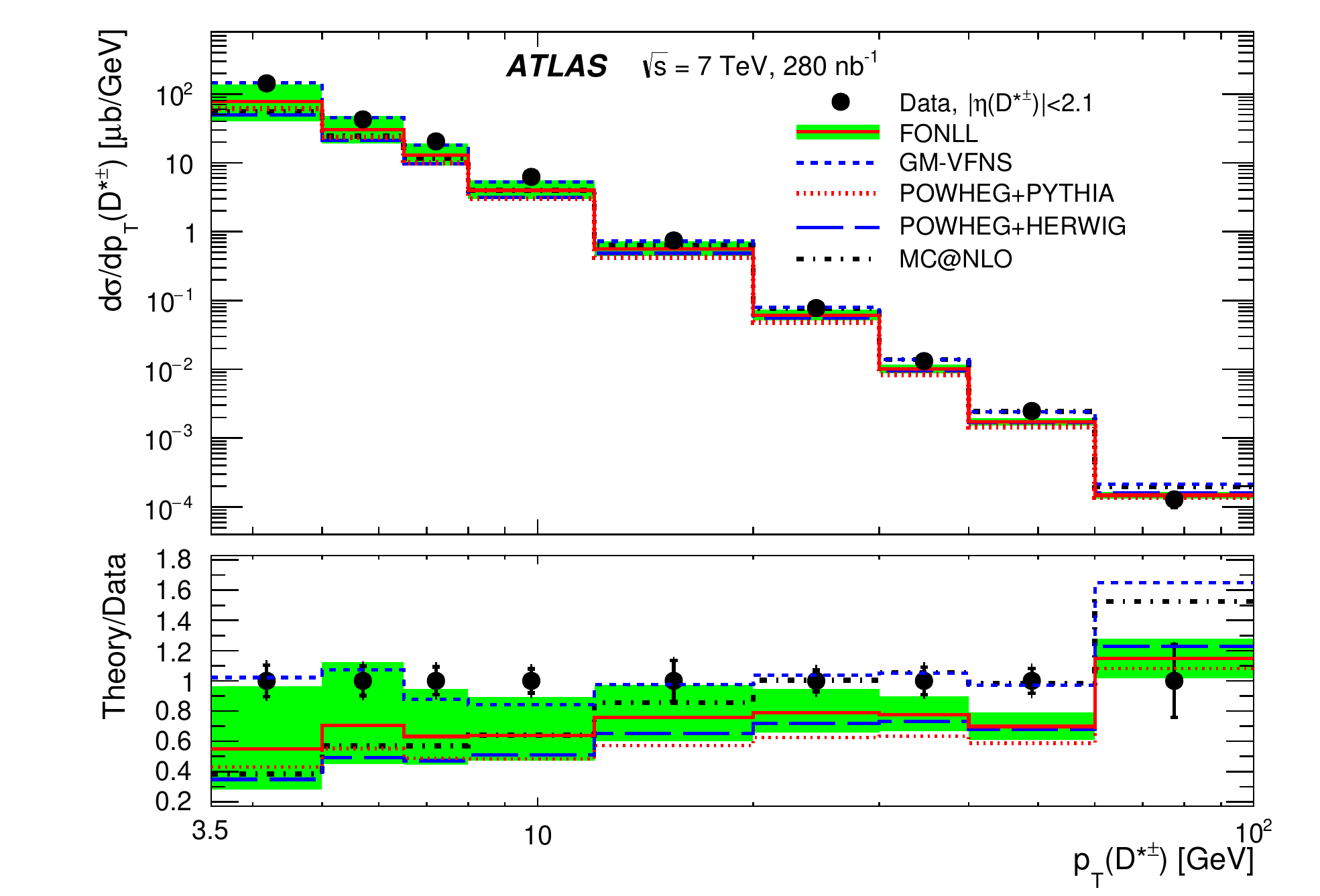}   
  \includegraphics[width=0.9\textwidth]{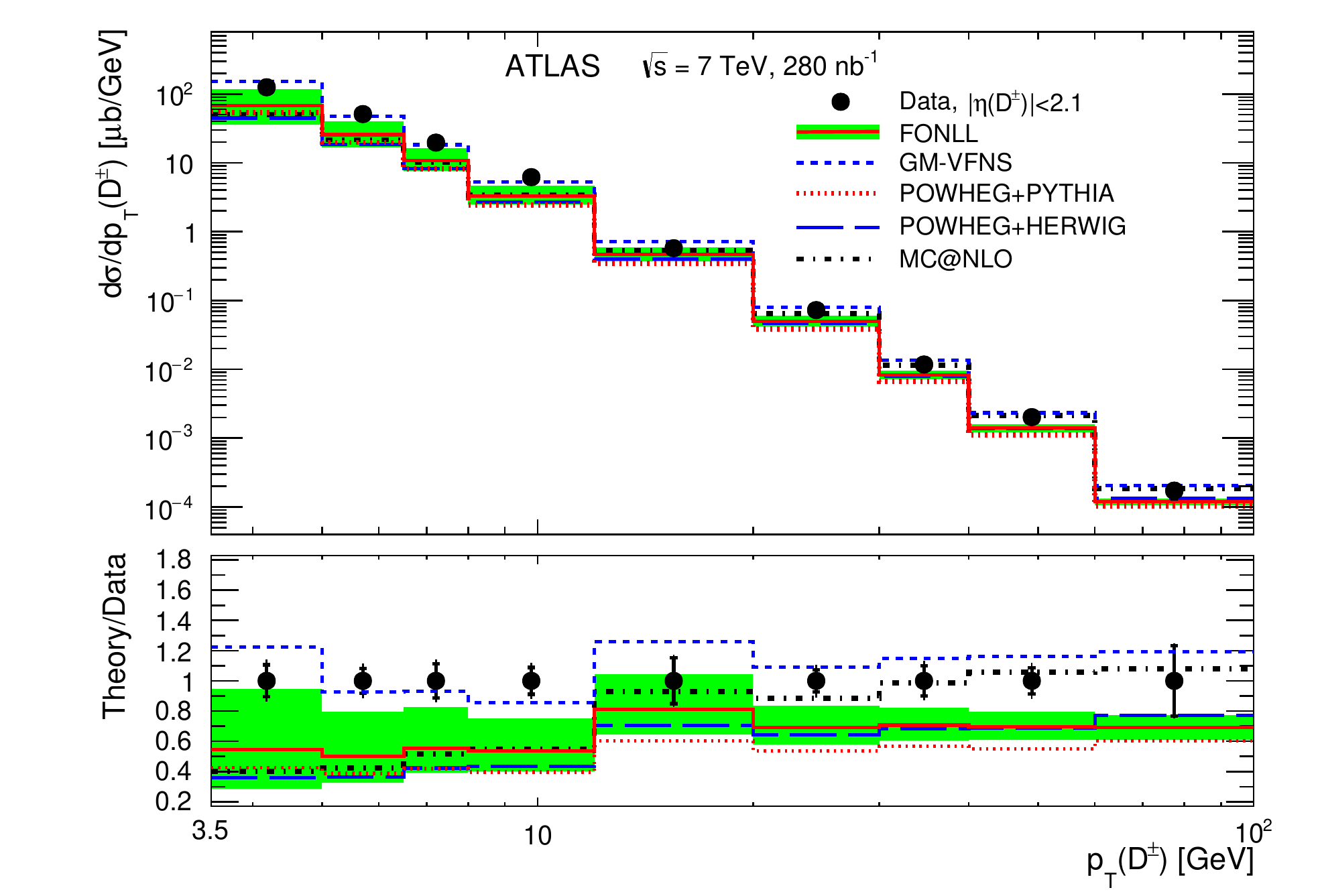}
  \caption{
Differential cross sections for $\dspm$ (top) and $\dcpm$ (bottom) mesons as a function
of $\pt$
for data (points)
compared to the NLO QCD calculations of
FONLL, POWHEG+PYTHIA, POWHEG+HERWIG, MC@NLO and GM-VFNS (histograms).
The data points are drawn in the bin centres.
The inner error bars show the statistical uncertainties
and the outer error bars
show the statistical and systematic
uncertainties added in quadrature.
Uncertainties linked with the luminosity measurement ($3.5\%$)
and branching fractions ($1.5\%$ and $2.1\%$
for $D^{*\pm}$ and $D^\pm$, respectively)
are not included in the shown systematic uncertainties.
The bands show the estimated theoretical
uncertainty of the FONLL calculation.
\label{fig:sig_dpt}}
\end{figure}

\begin{figure}[]
  \includegraphics[width=0.9\textwidth]{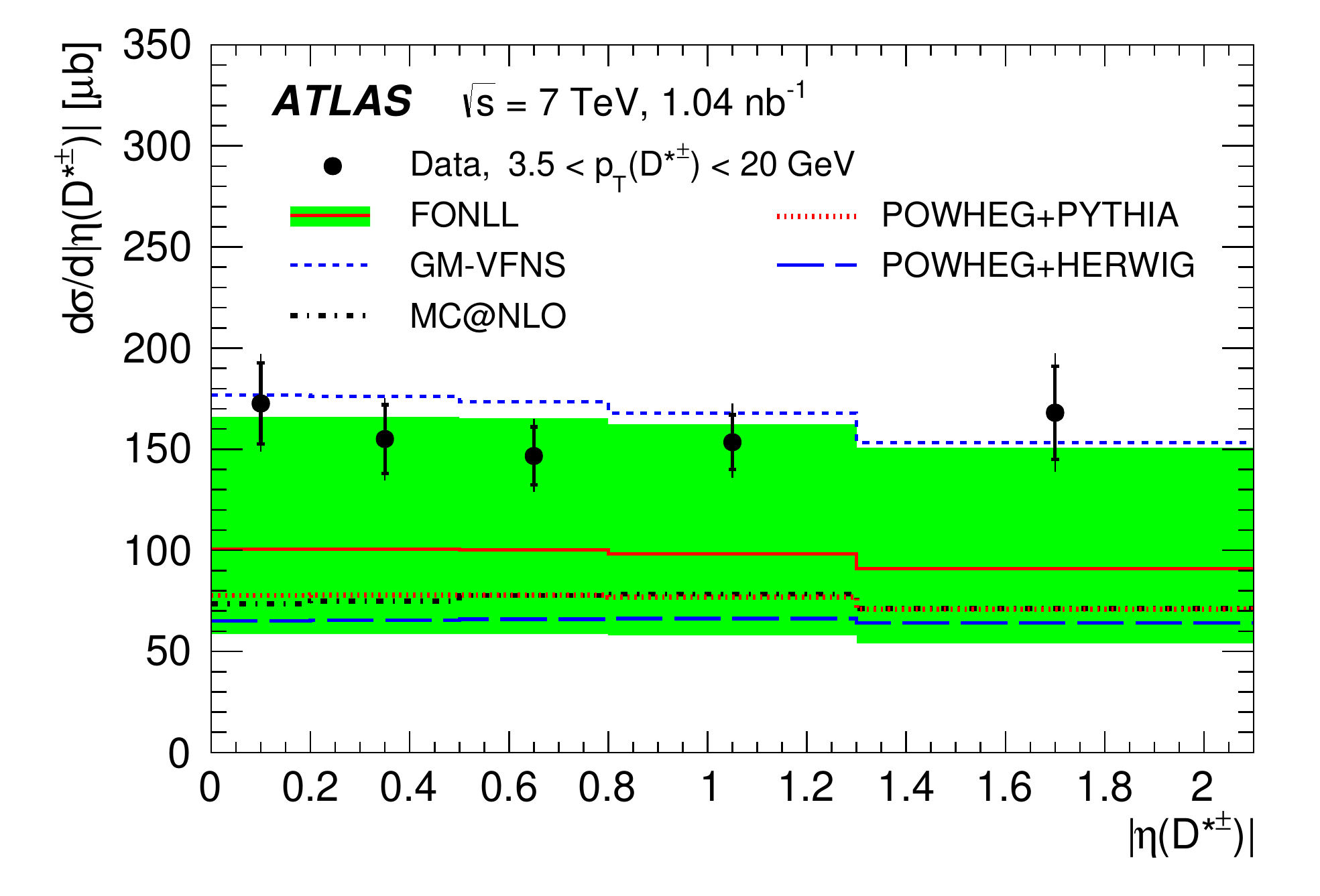}   
  \includegraphics[width=0.9\textwidth]{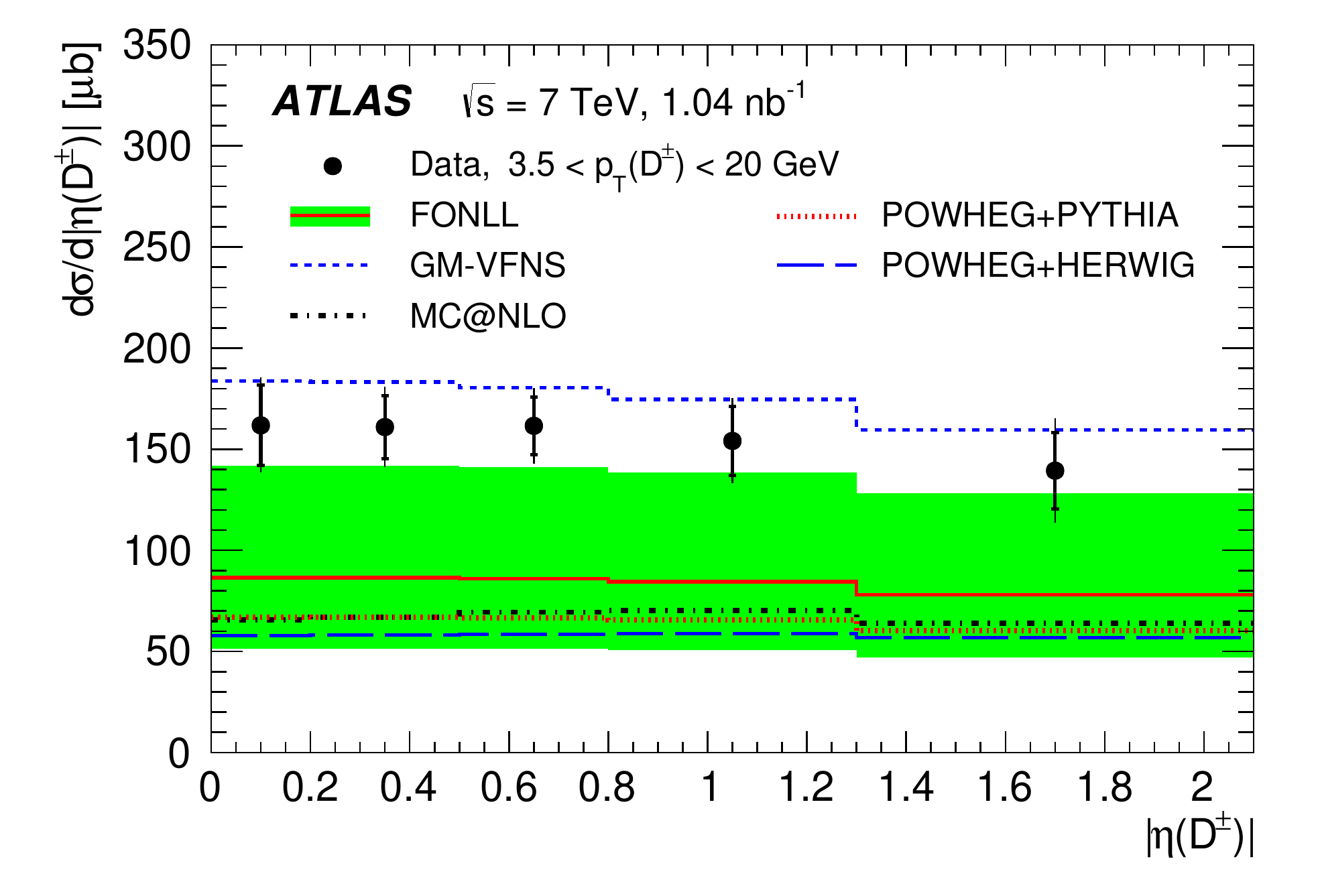}
  \caption{
Differential cross sections for $\dspm$ (top) and $\dcpm$ (bottom) mesons
with $3.5<\pt(D)<20\gev$
as a function
of $|\eta|$
for data (points)
compared to the NLO QCD calculations of
FONLL, POWHEG+PYTHIA, POWHEG+HERWIG, MC@NLO and GM-VFNS (histograms).
The data points are drawn in the bin centres.
The inner error bars show the statistical uncertainties
and the outer error bars
show the statistical and systematic
uncertainties added in quadrature.
Uncertainties linked with the luminosity measurement ($3.5\%$)
and branching fractions ($1.5\%$ and $2.1\%$
for $D^{*\pm}$ and $D^\pm$, respectively)
are not included in the shown systematic uncertainties.
The bands show the estimated theoretical
uncertainty of the FONLL calculation.
\label{fig:sig_detalpt}}
\end{figure}

\begin{figure}[]
  \includegraphics[width=0.9\textwidth]{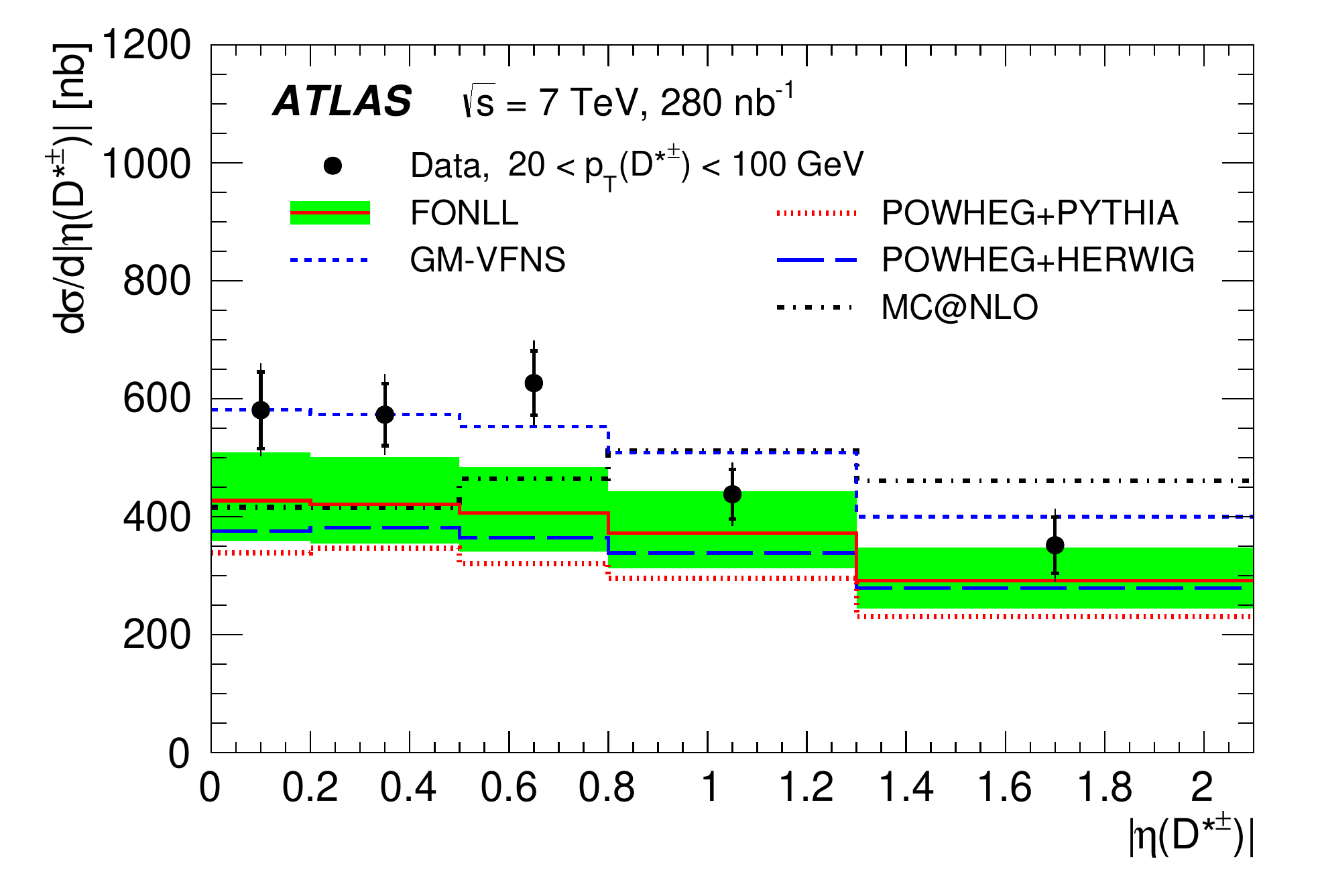}   
  \includegraphics[width=0.9\textwidth]{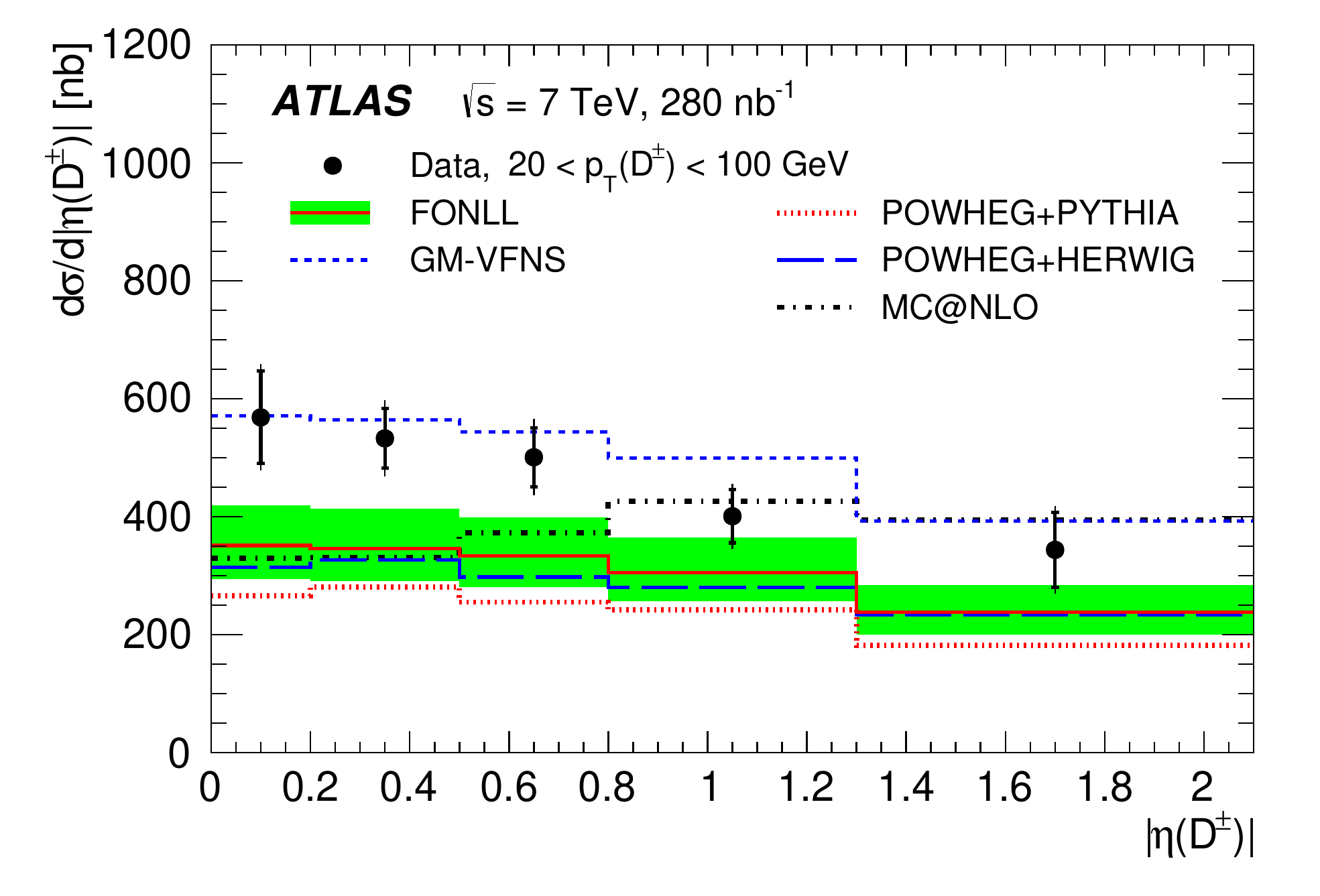}
  \caption{
Differential cross sections for $\dspm$ (top) and $\dcpm$ (bottom) mesons
with $20<\pt(D)<100\gev$
as a function
of $|\eta|$
for data (points)
compared to the NLO QCD calculations of
FONLL, POWHEG+PYTHIA, POWHEG+HERWIG, MC@NLO and GM-VFNS (histograms).
The data points are drawn in the bin centres.
The inner error bars show the statistical uncertainties
and the outer error bars
show the statistical and systematic
uncertainties added in quadrature.
Uncertainties linked with the luminosity measurement ($3.5\%$)
and branching fractions ($1.5\%$ and $2.1\%$
for $D^{*\pm}$ and $D^\pm$, respectively)
are not included in the shown systematic uncertainties.
The bands show the estimated theoretical
uncertainty of the FONLL calculation.
\label{fig:sig_detahpt}}
\end{figure}

The differential cross sections $\dr\sigma/\dr\pt$ and $\dr\sigma/\dr|\eta|$
for $\dspm$ and $\dcpm$ production
are shown in Tables~\ref{tab:xsdpt}-\ref{tab:xsdeta2}
and compared in
Figs.~\ref{fig:sig_dpt}-\ref{fig:sig_detahpt}
with the NLO QCD predictions.
The FONLL, MC@NLO and POWHEG predictions are generally below the data.
They are consistent with the data in the measured
$\pt(D)$ and $|\eta(D)|$ ranges 
within the large theoretical uncertainties.
The FONLL and POWHEG predictions reproduce shapes of the data distributions.
The $\pt$ shape of the MC@NLO prediction is harder than that for the data.
The $|\eta|$ shape of the MC@NLO prediction
in the high-$\pt$ range differs from the data
and
all other predictions.
The GM-VFNS predictions agree with data in both shape and normalisation.

\section{\label{secExtrap}Extrapolation to the full kinematic phase space}

The visible kinematic range covers
only a small fraction of produced charmed mesons.
To get some insight into the general properties
of charm
production and 
hadronisation at the LHC,
the visible low-$\pt$ $D$ cross sections are extrapolated to
the cross sections
in the full kinematic phase space
after subtracting the cross-section fractions
originating from beauty production.
Assuming the validity of the QCD NLO calculations
and QCD factorisation in the whole phase space,
the extrapolation factors are calculated
as ratios of the total NLO cross sections of $D$ mesons
produced in charm hadronisation, $\scct(D)$,
to those in the visible kinematic range.
The extrapolation factors from
the visible low-$\pt$
$\dsp$, $\dc$ and $\dssp$ cross sections
to the full kinematic phase space
are of the order $12$--$14$.

The extrapolated $D$ cross sections
are used to calculate
the total cross section of charm production
in $pp$ collisions at $\sqrt{s}=7\,$TeV,
and two charm fragmentation ratios:
the strangeness-suppression factor
in charm fragmentation and
the fraction of charged non-strange $D$ mesons
produced in a vector state.
The GM-VFNS calculations cannot be used for extrapolation
because they originate from the massless scheme.
For estimation of the total cross section of charm production,
the extrapolation is performed with the FONLL calculations.
However, as the FONLL calculations are not available for $\dssp$
production and do not include such a sophisticated fragmentation scheme
as PYTHIA,
the extrapolation for extraction of the
charm fragmentation ratios is performed with
the POWHEG+PYTHIA calculations.

The results obtained by extrapolating
the visible high-$\pt$ $D$ cross sections
agree with the results presented,
but have larger extrapolation uncertainties.

\subsection{\label{secTot}Total charm production cross section}

To calculate the total cross section of charm production,
the total production cross section of a given $D$ meson
should be divided by twice the value
of the corresponding charm fragmentation fraction from Table~\ref{tab:ftod}.
The weighted mean of the two values
calculated from $\dspm$ and $\dcpm$  cross sections is
$$\scct=8.6\pm0.3\,({\rm stat})\pm0.7\,({\rm syst})\pm 0.3\,({\rm lum})\pm 0.2\,({\rm ff})^{+3.8}_{-3.4}\,({\rm extr})\,{\rm mb}~~~~({\rm ATLAS})\,,$$
where the fourth uncertainty is due to the
uncertainty
of the fragmentation fractions
and the last uncertainty is due to
the extrapolation procedure.
The extrapolation uncertainty
is determined
by adding in quadrature the changes in results originating from
all sources of the FONLL theoretical uncertainty (Section~\ref{secNLO}).
The uncertainties in the charmed meson decay branching fractions,
which are common to the measured cross sections and
fragmentation fractions,
do not affect the calculation
of the total cross section of charm production.

The calculated total cross section of charm production
can be compared with a similar calculation performed
by the ALICE experiment~\cite{ALICE2Ds}:
$$\scct=8.5\pm0.5\,({\rm stat})^{+1.0}_{-2.4}\,({\rm syst})\pm0.3\,({\rm lum)}\pm0.2\,({\rm ff})^{+5.0}_{-0.4}({\rm extr})\,{\rm mb}~~~~({\rm ALICE})\,.$$
The ATLAS and ALICE estimates of the total charm production cross section at LHC are in good agreement.
Both estimations are performed using extrapolations outside the visible kinematic ranges with
analogous FONLL calculations.
The different extrapolation uncertainties of the two estimations are due
to different visible kinematic ranges. ATLAS extrapolates from the kinematic
range $3.5<\pt(D)<20\gev$ and $|\eta(D)|<2.1$,
while the ALICE visible kinematic range is
$1<\pt(D)<24\gev$ and $|y(D)|<0.5$.

\subsection{\label{secFragm}Charm fragmentation ratios}

The total cross sections for $D$ production
are used to calculate two fragmentation ratios
for charged charmed mesons:
the strangeness-suppression factor,
$\gamma_{s/d}$, and the fraction of charged non-strange $D$ mesons
produced in a vector state, $P_{\rm v}^d$.
The strangeness-suppression factor
is calculated as the ratio of the $\scct(D_s^{+})$
to the sum of $\scct(D^{*+})$
and that part of $\scct(D^{+})$ which
does not originate from $\dsp$ decays:
$$\gamma_{s/d}=\frac{\scct(D_s^{+})}{\scct(D^{*+})+\scct(D^{+})-\scct(D^{*+})\cdot(1-\br)}=\frac{\scct(D_s^{+})}{\scct(D^{+})+\scct(D^{*+})\cdot\br}\,,$$
where $\br=0.677\pm0.005\,$~\cite{pdg2014} is the branching fraction
of the $D^{*+}\rightarrow D^0\pi^+$ decay. 
The fraction of charged non-strange $D$ mesons produced in a vector state
is calculated as the ratio of $\scct(D^{*+})$
to the sum of $\scct(D^{*+})$
and that part of $\scct(D^{+})$ which
does not originate from $\dsp$ decays:
$$P_{\rm v}^d=\frac{\scct(D^{*+})}{\scct(D^{*+})+\scct(D^{+})-\scct(D^{*+})\cdot(1-\br)}=\frac{\scct(D^{*+})}{\scct(D^{+})+\scct(D^{*+})\cdot\br}\,.$$

The large extrapolation uncertainties,
which affect the extrapolated cross sections, 
are expected to nearly cancel out
in the ratios.
However, the calculations of the ratios are affected by details
of the fragmentation simulation.
To determine the extrapolation uncertainties,
the following variations of the PYTHIA fragmentation,
in addition to the POWHEG+PYTHIA theoretical uncertainty (Section~\ref{secNLO}),
are considered:
\begin{itemize}
\item{the Bowler fragmentation function parameter $r_c$ is varied from
the predicted value of $1$ to $0.5$; the $a$ and $b$ parameters of the Lund
symmetric function are varied by $\pm20\%$ around their default values;
}
\item{the PYTHIA parameter for the strangeness suppression is taken to be $0.3\pm0.1$;
}
\item{the PYTHIA parameter for the fraction of the lowest-mass charmed mesons produced
in a vector state is taken to be $0.6\pm0.1$;
}
\item{the PYTHIA parameters for production rates of the excited charmed and charmed-strange mesons
are varied by $\pm50\%$ around the central values tuned to reproduce
the measured fractions of $c$ quarks hadronising into
$D^0_1$, $D^{*0}_2$ or $D^+_{s1}$~\cite{ZEUS_dexc}.
}
\end{itemize}

Using the extrapolated cross sections, the strangeness-suppression factor
and
the fraction $P_{\rm v}^d$ are
$$\gamma_{s/d}=0.26\pm0.05\,({\rm stat})\pm0.02\,({\rm syst})\pm 0.02\,({\rm br})\pm0.01\,({\rm extr})\,,$$
$$P_{\rm v}^d=0.56\pm0.03\,({\rm stat})\pm0.01\,({\rm syst})\pm 0.01\,({\rm br})\pm0.02\,({\rm extr})\,.$$
The measured $P_{\rm v}^d$ fraction is smaller
than the naive spin-counting prediction of $0.75$,
suggesting the charm-quark mass is not large enough
to ensure a precise description of charm fragmentation
by heavy-quark effective theory~\cite{david}.
The predictions of the thermodynamical approach~\cite{becattini}
and the string fragmentation approach~\cite{pei},
which both predict $2/3$ for the fraction,
are closer to, but still above,
the measured value.

The measured charm fragmentation ratios agree with those measured
by ALICE~\cite{ALICE7Ds,ALICE7Dss}
and those measured at the HERA collider
in $e^\pm p$ collisions~\cite{H1ffdis,ZEUSffH1,ZEUSffdis,ZEUSffH2}.
They can also be compared with results obtained
in $e^+e^-$ annihilations at LEP.
The LEP fragmentation ratios are calculated
using the fragmentation fractions from Table~\ref{tab:ftod}:
$$\gamma^{\rm LEP}_{s/d}=\frac{\fcdss}{\fcdc+\fcds\cdot\br}=0.24\pm0.02\pm0.01\,({\rm br})\,,$$
$$P^{\rm LEP}_{\rm v}=\frac{\fcds}{\fcdc+\fcds\cdot\br}=0.61\pm0.02\pm0.01\,({\rm br})\,,$$
where the first uncertainties are the combined statistical and systematic
uncertainties of the LEP measurements and
the second uncertainties originate from
uncertainties
of the relevant branching fractions.
The measurements agree within experimental uncertainties,
in agreement with the hypothesis of charm fragmentation universality.

\section{Summary}

The production of $\dspm$, $~\dcpm$ and $~\dsspm$ charmed mesons
has been measured
in the kinematic region
$3.5<\pt(D)<100\gev$ and $|\eta(D)|<2.1$
with the ATLAS detector in $pp$ collisions at $\sqrt{s}=7\,$TeV
at the LHC,
using an integrated luminosity of up to $280\,$nb$^{-1}$.
The differential cross sections $\dr\sigma/\dr\pt$ and $\dr\sigma/\dr |\eta|$
for $\dspm$ and $\dcpm$ production
have been determined and compared
with a number of NLO QCD predictions.
The FONLL~\cite{fonll1,fonll2,fonll3,fonllweb},
MC@NLO~\cite{mcatnloMETHOD,mcatnloHQ}
and
POWHEG ~\cite{powhegMETHOD,powhegHQ}
predictions are generally below the data.
They are consistent with the data in normalisation
within the large theoretical uncertainties.
The FONLL and POWHEG predictions reproduce the shapes of the data distributions
while the MC@NLO predictions show deviations from the shapes in the data.
The GM-VFNS~\cite{gmvfns1,gmvfns2,gmvfns3} predictions agree with data
in both shape and normalisation.

Using the visible $D$ cross sections
and an extrapolation to the full kinematic phase space,
the strangeness-suppression factor
in charm fragmentation,
the fraction of charged non-strange $D$ mesons
produced in a vector state,
and the total cross section of charm production
in $pp$ collisions at $\sqrt{s}=7\,$TeV
have been calculated.
The fragmentation ratios agree with those
obtained by the ALICE collaboration at the LHC, and those
measured in $e^+e^-$ annihilations at LEP and
in $e^\pm p$ collisions at HERA.
The total cross section of charm production
at $\sqrt{s}=7\,$TeV agree with the result of the ALICE collaboration.

\section*{Acknowledgements}

We thank CERN for the very successful operation of the LHC, as well as the
support staff from our institutions without whom ATLAS could not be
operated efficiently.

We acknowledge the support of ANPCyT, Argentina; YerPhI, Armenia; ARC, Australia; BMWFW and FWF, Austria; ANAS, Azerbaijan; SSTC, Belarus; CNPq and FAPESP, Brazil; NSERC, NRC and CFI, Canada; CERN; CONICYT, Chile; CAS, MOST and NSFC, China; COLCIENCIAS, Colombia; MSMT CR, MPO CR and VSC CR, Czech Republic; DNRF, DNSRC and Lundbeck Foundation, Denmark; IN2P3-CNRS, CEA-DSM/IRFU, France; GNSF, Georgia; BMBF, HGF, and MPG, Germany; GSRT, Greece; RGC, Hong Kong SAR, China; ISF, I-CORE and Benoziyo Center, Israel; INFN, Italy; MEXT and JSPS, Japan; CNRST, Morocco; FOM and NWO, Netherlands; RCN, Norway; MNiSW and NCN, Poland; FCT, Portugal; MNE/IFA, Romania; MES of Russia and NRC KI, Russian Federation; JINR; MESTD, Serbia; MSSR, Slovakia; ARRS and MIZ\v{S}, Slovenia; DST/NRF, South Africa; MINECO, Spain; SRC and Wallenberg Foundation, Sweden; SERI, SNSF and Cantons of Bern and Geneva, Switzerland; MOST, Taiwan; TAEK, Turkey; STFC, United Kingdom; DOE and NSF, United States of America. In addition, individual groups and members have received support from BCKDF, the Canada Council, CANARIE, CRC, Compute Canada, FQRNT, and the Ontario Innovation Trust, Canada; EPLANET, ERC, FP7, Horizon 2020 and Marie Skłodowska-Curie Actions, European Union; Investissements d'Avenir Labex and Idex, ANR, Region Auvergne and Fondation Partager le Savoir, France; DFG and AvH Foundation, Germany; Herakleitos, Thales and Aristeia programmes co-financed by EU-ESF and the Greek NSRF; BSF, GIF and Minerva, Israel; BRF, Norway; the Royal Society and Leverhulme Trust, United Kingdom.

The crucial computing support from all WLCG partners is acknowledged
gratefully, in particular from CERN and the ATLAS Tier-1 facilities at
TRIUMF (Canada), NDGF (Denmark, Norway, Sweden), CC-IN2P3 (France),
KIT/GridKA (Germany), INFN-CNAF (Italy), NL-T1 (Netherlands), PIC (Spain),
ASGC (Taiwan), RAL (UK) and BNL (USA) and in the Tier-2 facilities
worldwide.

\newpage
\begin{flushleft}
{\Large The ATLAS Collaboration}

\bigskip

G.~Aad$^\textrm{\scriptsize 85}$,
B.~Abbott$^\textrm{\scriptsize 113}$,
J.~Abdallah$^\textrm{\scriptsize 151}$,
O.~Abdinov$^\textrm{\scriptsize 11}$,
R.~Aben$^\textrm{\scriptsize 107}$,
M.~Abolins$^\textrm{\scriptsize 90}$,
O.S.~AbouZeid$^\textrm{\scriptsize 158}$,
H.~Abramowicz$^\textrm{\scriptsize 153}$,
H.~Abreu$^\textrm{\scriptsize 152}$,
R.~Abreu$^\textrm{\scriptsize 30}$
% ...
\bigskip
\\
% ...
$^{11}$ Institute of Physics, Azerbaijan Academy of Sciences, Baku, Azerbaijan\\
% ...
$^{30}$ CERN, Geneva, Switzerland\\
% ...
$^{85}$ CPPM, Aix-Marseille Universit{\'e} and CNRS/IN2P3, Marseille, France\\
% ...
$^{90}$ Department of Physics and Astronomy, Michigan State University, East Lansing MI, United States of America\\
% ...
$^{107}$ Nikhef National Institute for Subatomic Physics and University of Amsterdam, Amsterdam, Netherlands\\
% ...
$^{113}$ Homer L. Dodge Department of Physics and Astronomy, University of Oklahoma, Norman OK, United States of America\\
% ...
$^{151}$ Institute of Physics, Academia Sinica, Taipei, Taiwan\\
$^{152}$ Department of Physics, Technion: Israel Institute of Technology, Haifa, Israel\\
$^{153}$ Raymond and Beverly Sackler School of Physics and Astronomy, Tel Aviv University, Tel Aviv, Israel\\
% ...
$^{158}$ Department of Physics, University of Toronto, Toronto ON, Canada\\
% ...
\end{flushleft}


\end{document}